\documentstyle[aps,12pt,epsfig]{revtex}

\newcommand{\br}{{\bf r}}
\newcommand{\bx}{{\bf x}}
\newcommand{\bv}{{\bf v}}
\newcommand{\hr}{\br}
\newcommand{\hdr}{\delta{\br}}
\newcommand{\bdx}{\delta{\bx}}
\newcommand{\bdr}{\bdx}
\newcommand{\BE}{\begin{equation}}
\newcommand{\EE}{\end{equation}}
\newcommand{\BA}{\begin{eqnarray}}
\newcommand{\EA}{\end{eqnarray}}

\begin{document}

%\title{
\begin{center}{\huge
Spatial Patterns in Chemically and Biologically Reacting
Flows}\footnotetext{To appear on the Proceedings of the 2001
ISSAOS School on {\sl Chaos in Geophysical Flows}. }
\vskip 1cm
%\author{

{\Large Emilio Hern\'{a}ndez-Garc\'{\i}a$^1$\footnote{web page:
{\tt http://www.imedea.uib.es/\~{ }emilio/}} , Crist\'{o}bal
L\'{o}pez$^2$, and Zolt\'{a}n Neufeld$^3$}

\vskip 0.5cm

{\large \baselineskip 18pt $^1$IMEDEA (CSIC-UIB) Instituto
Mediterr\'{a}neo de Estudios Avanzados, Campus Universitat de les
Illes Balears\\ E-07071 Palma de Mallorca, Spain
\\$^2$ Dipartimento di Fisica, Universit\`{a} di Roma `La Sapienza',
P.le A. Moro 2, I-00185, Roma, Italy
\\$^3$ Department of Applied Mathematics and Theoretical Physics,
University of Cambridge, Silver Street, Cambridge CB3 9EW, UK }

%\maketitle
%\vskip 0.5cm
%ISSAOS, L'Aquila, September 2001
\end{center}
%\maketitle
%\vskip 1cm

%\baselineskip 24pt
\section{Introduction}
\label{sec:intro}
Reacting complex flows are ubiquitous in the atmosphere, the
oceans, and the interior of our planet:  Stratospheric ozone
chemistry is perhaps the most well known example of chemical
dynamics occurring in a stirred geophysical fluid. But the fate of
many other reacting tracers is also influenced by mesoscale eddies
in the ocean, or by convection in the Earth mantle, to name just a
few cases. Biological population dynamics is also a kind of
``chemical reaction". The interactions of nutrients and plankton
species in the sea are the first steps in the ocean food chain,
and they are among the most important ingredients for the
understanding the CO$_2$ interchange between atmosphere and ocean.

Tracers stirred by fluid motion are known to develop strong
inhomogeneities, usually in the form of filamental features,
arising from a kind of variance cascade from the forcing scale
towards smaller scales. These structures are poorly resolved in
global atmospheric and, especially, in oceanic models. They
provide however sensitive mixing mechanisms and they are in some
sense ``catalists" \cite{karoly99} that enhance the chemical or
biological activity occurring in geophysical flows. We mention as
examples the increase of ozone depletion produced by the
filamental structure of chlorine filaments \cite{edouard}, or the
increase in biological production in small-scale ocean structures
\cite{mahadevan}.

In this set of Lectures, we will present some theoretical ideas
for the study of structures arising in advected reacting tracers.
The emphasis will be in the modifications that chemical or
biological activity introduces in reactive patterns with respect
to the ones appearing in passively advected substances.

Examples from plankton populations dynamics will be used through
these Lectures. It should be mentioned, however, that if one looks
into the literature, the diversity of observations related to
plankton distributions is so great that some authors, rather
pessimistically, argue that ``{\sl ... One should expect no
general results that describe in all (or most) cases how 'the
biology' modifies a spatial pattern that has arisen solely from
the dispersal of species ...}" \cite{PowellOkubo}. We use here the
biological experimental observations more as a motivation and
illustration, than as a detailed check of a particular theory. Our
aim is to point out simple and robust mechanisms that can be
useful in classifying the large number of observations into a more
reduced set of different scenarios. Later quantitative test would
need synoptic measurements of the biological variables {\sl and}
of the hydrodynamic flows. This would be certainly more accurately
done in laboratory chemical experiments than in open sea biology.
Most of the results presented here are of direct application to
chemical reactions of the decaying or of the excitable type.
Nevertheless, we mention that comparison of some of our
theoretical ideas with simultaneous satellite data of both a
reactive tracer (sea temperature in interaction with air
temperature) and hydrodynamic flow has been very recently
initiated \cite{Abraham2}.

In Section \ref{sec:plankton} we will briefly review some aspects
of plankton dynamics that will be useful in the following. The
paradigm of chaotic advection, i.e. the chaotic transport of fluid
parcels by smooth large-scale flows, will be the one more
frequently used here to discuss flow effects, and it is described
in Section \ref{sec:chaoadvect}. Sections \ref{sec:dps} and
\ref{sec:excitable} describe results obtained for two rather
general classes of chemical or biological reactions in chaotic
flows: the first-order reaction, or linearly decaying tracer, and
reactions of the excitable type. Finally (Sect.
\ref{sec:individual}), some warnings will be given about the
difficulties in modeling discrete individuals (such as planktonic
organisms) in terms of continuous concentration fields.

\section{Plankton, patches, and blooms}
\label{sec:plankton}

\subsection{Background}
\label{subsec:background}

Plankton is the generic name given to a huge variety of aquatic
organisms, both from marine and from freshwater environments,
comprising from viruses with less than 200 nm in size, to
crustaceans or even fish larvae, with sizes above the centimeter.
What they have in common, and distinguishes them from other major
group called {\sl nekton}, is their inability to overcome the
major ambient hydrodynamic currents. In this sense, they are
strongly influenced by the existent {\sl hydrodynamic
weather}\cite{catalan}. It is often assumed that they are
passively transported by the marine currents. This is probably
true for the smallest species, but it should be said that most of
these organisms present flagellae and other natatory organs that
give to them some mobility, especially in the vertical direction.
In addition to interactions with the hydrodynamic environment
different plankton species interact among them in a variety of
ways, such as predator-prey interaction, parasite-host
relationship, competition for resources, etc. The biological
interaction between species and with the nutrient substances
dissolved in water is formally equivalent to a chemical reaction
dynamics.

Any attempt to overview here some major theme in plankton biology
and its interaction with hydrodynamics would result very partial
and incomplete, and we refer the interested reader to some
monographs\cite{BarnesHughes,MannLazier} that turn out to be very
readable. We just mention that a major division between plankton
species arises between the ones that are able to assimilate
inorganic material by photosynthesis, the {\sl phytoplankton}, and
the species that necessarily graze on the above, the {\sl
zooplankton}. In the following we will briefly describe just two
phenomena, {\sl plankton patchiness} and {\sl plankton blooms},
extremely stimulating from the perspective of cross-disciplinary
research, and for which some ideas from Nonlinear Science may be
of relevance.

Plankton patchiness \cite{mackas} is the term used to label the
large inhomogeneity observed in plankton distributions. Earlier
observations from land and from ships reported water masses, of
sizes around say 10 km, distinctly colored by the presence of
different species of planktonic organisms. The search for
biological or physical mechanisms singling out a characteristic
patch size, despite mixing with the surrounding water stimulated
the first theoretical approaches\cite{kiSs,KIsS}. Modern satellite
observations\cite{seawifs} reveal that the range of sizes of
plankton structures is much broader. The largest features,
thousand of kilometers wide, are associated with major ocean gyres
and currents. Upwellings of nutrient-rich water from deeper ocean
layers, and river discharges, fertilize the upper layer of the
ocean and induce associated plankton structures. In the mesoscale
range, fronts and eddies are seen as major players, both in
bringing to the surface nutrients from lower layers, and in
providing horizontal transport and stirring processes.

\begin{center}
\begin{figure}[t]
\epsfig{file=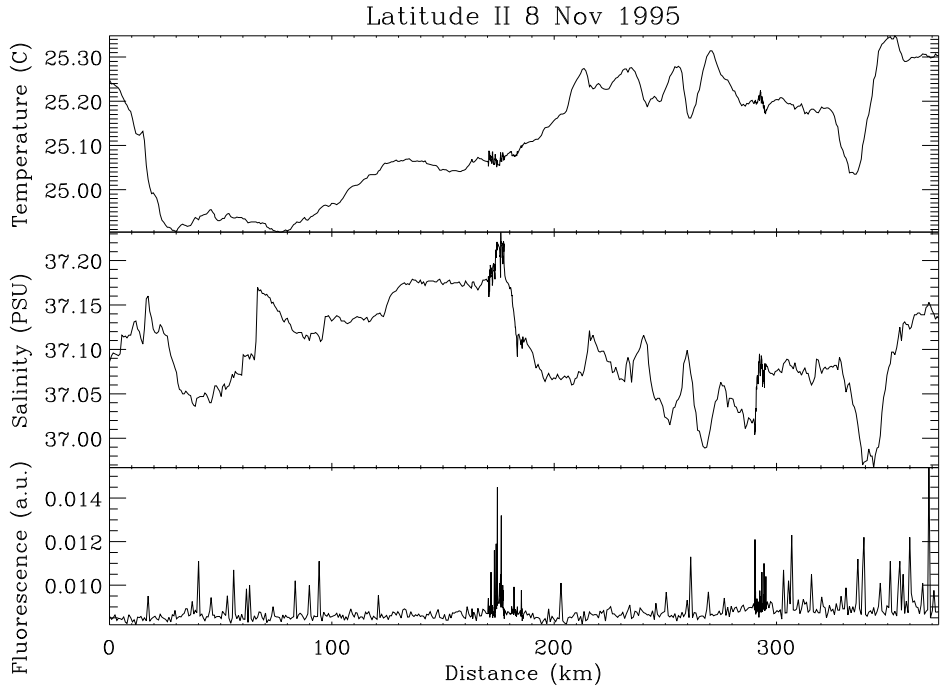,width=0.8\textwidth}
\vskip 0.5cm \nopagebreak
\caption{\protect\footnotesize
Temperature (in degrees C), salinity (in Practical Salinity Units)
and induced fluorescence (in arbitrary units) measured at 5 m
depth along the path of the research vessel BIO Hesp\'{e}rides,
across the southern Atlantic tropical gyre during the Latitude II
campaign\protect\cite{agusti}. Fluorescence identifies chlorophyll
and is thus a proxy for phytoplankton concentration. The starting
point of the transect shown was at 31.65 W, 16.48 S, and the final
point at 32.83 W, 19.43 S. }
\label{fig:transect}
\end{figure}
\end{center}

A closer look at the inhomogeneities in plankton distributions can
be performed by analyzing water samples along a ship trajectory
(or {\sl transect}). Higher spatial resolution can then be
attained and the result is an extremely intermittent distribution
(Fig.~\ref{fig:transect}, lower panel). At this point, it is
pertinent to ask about the relevance of the fact that plankton is
a living substance, or if perhaps the same kind of distributions
can be found for non-biological tracers. A partial answer to this
question is given by Fig.~\ref{fig:transect}, in which
distributions of physically controlled tracers (temperature and
salinity) are compared with the phytoplankton concentration,
indicated by water induced fluorescence, at a resolution of about
half a kilometer. Although correlations can be appreciated,
evident differences are seen between the distributions of both
kinds of tracers. A brief walk on the literature shows that the
distributions may look very different form place to place in the
ocean, and that their characteristics may also change in
time\cite{timevar}. In general one can say that they will be
affected by both the characteristics of the fluid flow that
transports them, and by the chemical or biological interactions in
which they are involved.

The irregular nature of the observed distributions immediately
suggest characterization in terms of scaling exponents. This has
traditionally been done in terms of the power spectrum of
concentration fluctuations\cite{DenmanPlatt}, since a comparison
with Kolmogorov-like turbulence theories was then direct. In
situations in which the phytoplankton concentration spectrum
displayed a large and well defined power-law regime (which is not
the case of the data in Fig.~\ref{fig:transect}) it decayed as
$k^{-\beta}$, with $k$ the wavenumber. The scaling exponent
$\beta$ results to be larger than one. We prefer to discuss the
scaling properties of concentration fluctuations in terms of
structure functions, since a more complete characterization of
intermittency properties can be done. These quantities, and their
relation with the power spectrum, are introduced in Sect.
\ref{sec:dps}.

A second kind of inhomogeneity found in plankton distributions is
{\sl temporal} inhomogeneity associated to plankton blooms.
Seasonal blooms are episodes of rapid phytoplankton growth that
occur in mid-latitude seas at the beginning of spring. This is
followed by zooplankton (and also bacterial) growth that consumes
the available phytoplankton until concentrations return back to
the initial low level. The whole cycle finishes with the summer.
Sometimes a smaller version of the bloom repeats in autumn. The
mechanism for seasonal blooms seems to be understood, at least
qualitatively: the beginning of ocean stratification in spring
reduces the depth of the layer in which phytoplankton is mixed by
turbulent motion, the mixed layer, so that these
photosynthesis-capable organisms spend more time exposed to solar
light. Fast growth then occurs. The subsequent growth of
predators, and the consumption of nutrients in the upper layer,
ends the bloom. Destratification, with the destruction of the
seasonal thermocline in autumn, brings again to the surface some
nutrients from the deeper layers, so that phytoplankton activity
may increase again.

There is another kind of blooms that occur more localized in
space, and not with a seasonal periodicity, but occasionally and
associated to local warming or local increase of nutrients in the
water. They are called sometimes {\sl red tides} since the large
amount of organisms in water may give to it some distinct
coloration. Time scales for the initial development of the bloom
are in the range of days, and the end usually arrives in a time of
the order of weeks or month.

\subsection{Models}
\label{subsec:models}

Modeling the above spatial and temporal evolutions of planktonic
populations is a challenge since many physical and biological
processes are involved, some of them poorly understood. A variety
of levels of description are available, from detailed models based
on individual organism behavior, to mean field approaches that
consider the whole sea as an homogeneous soup without spatial
features. Another aspect of modeling that should be addressed is
the degree of aggregation at which groups of species are
represented by the same collective variables. It should be said
that these starting points, the election of the adequate levels of
description, and the relationship between different degrees of
detail, is an important subject by itself, and its complexity is
just beginning to be grasped\cite{scales1,scales2}. More on this
in Sect. \ref{sec:individual}.

\begin{center}
\begin{figure}[t]
\epsfig{file=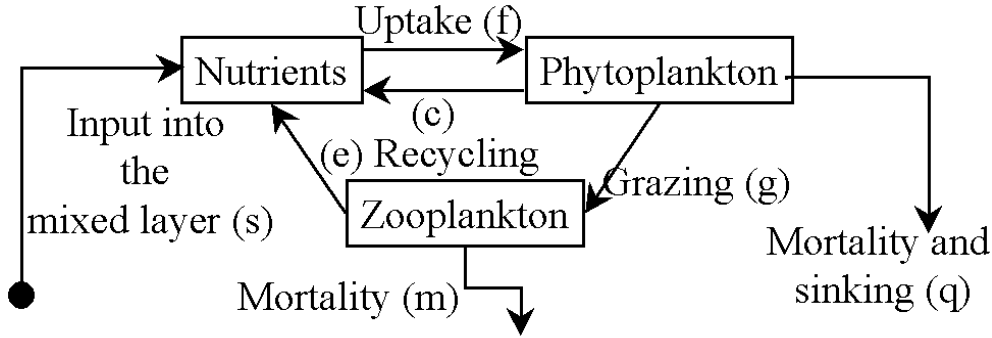,width=0.8\textwidth}
\nopagebreak
\caption{\protect\footnotesize The processes in the NPZ models. The letter in parenthesis identifies
the process with the term representing it in Eqs. (\ref{NPZ}).}
\label{fig:NPZ}
\end{figure}
\end{center}

A class of models simple enough to allow some explicit analysis,
but still retaining some of the complexities of real-life trophic
interactions is the one known as the NPZ class of models (from
Nutrients, Phytoplankton and Zooplankton). We describe it here for
illustration, and because it will be of use in some of the results
presented later. The processes included within this framework are
displayed in Fig. \ref{fig:NPZ}, and its mathematical modeling is
as follows\cite{Brindley94b,Edwards01}: in each portion of well
mixed fluid, the time evolution of the amount of nutrients ($N$),
phytoplankton ($P$), and zooplankton ($Z$) is ruled by
\BA
\frac{dN}{dt} &=& F_N(N,P,Z) = s(N_0 -N) -f(N) P +c P +a g(P)Z +e Z \nonumber
\\
\frac{dP}{dt} &=& F_P(N,P,Z) = f(N) P - q P - g(P)Z  \nonumber  \\
\frac{dZ}{dt} &=& F_Z(N,P,Z) = k g(P)Z - m Z^r   \ \ \ .
\label{NPZ}
\EA
The nutrient input $N_0$ affects the nutrient concentration in the
mixed region, and this influence is transferred to phytoplankton
by the second equation, and to zooplankton by the third. Finally,
zooplankton either dies or is consumed by higher organisms, which
leads to the last term in the third equation. The exponent $r$ in
this last term is usually taken to be either $r=1$ or $2$,
depending on the dominant kind of predation on zooplankton one
wants to model.

For the description of the spatial dependence, the framework of
{\sl Advection-Reaction-Diffusion} equations is frequently used.
They are partial differential equations in which the organisms and
their nutrients are described by continuous distributions
subjected to advection by a velocity field $\bv(\bx,t)$, and to
diffusion, in addition to the interactions (\ref{NPZ}). The
general form of the equations is
\BE
\frac{\partial C_i}{\partial t} +\nabla \cdot \left( \bv C_i \right) =
F_i (N,P,Z) + D_i \nabla^2 C_i \ \ .
\label{ard}
\EE
The index $i$ takes three values, so that $F_i$ is either $F_N$,
$F_P$, or $F_Z$, and the variables $C_i$ are either $N$, $P$, or
$Z$ (and now these symbols denote {\sl concentrations} per unit
volume). If the flow is incompressible ($\nabla \cdot \bv =0$)
then the second term in the left-hand-side can be written as $\bv
\cdot \nabla C_i$.  At scales large enough, so that the effects of
diffusion can be neglected, Eq. (\ref{ard}) admits a Lagrangian
representation that in the incompressible-flow case coincides with
the system in Eq. (\ref{NPZ}), where now $N$, $P$, and $Z$ are
either the concentration or the amount of substance contained in a
fluid element that moves with the flow velocity. The trajectory
$\br(t)$ of this fluid parcel thus satisfies
\BE
\frac{d\br}{dt} = \bv (\br,t) \ \ .
\label{trajectory}
\EE
It may seem that Eqs. (\ref{NPZ}) and (\ref{trajectory}) represent
uncoupled dynamical systems. This would be the case if none of the
parameters in (\ref{NPZ}) is spatially dependent. But if one of
them varies from point to point, it should be evaluated at the
position of the fluid element at time $t$, that is, at $\br(t)$.
This couples the {\sl chemical} and the {\sl transport} dynamical
systems. For example, the parameter $N_0$ may represent an
inhomogeneous nutrient input $N_0(\bx)$ into the upper ocean mixed
layer by a localized upwelling, and thus the fluid element
following the trajectory $\br(t)$ would experience a
time-dependent input $N_0(\bx=\br(t))$.

The model is still largely undefined until one specifies the
parameters and, particularly, the response functions $f$ and $g$,
which really contain information about the species interaction. A
rather common form for the function $f$, experimentally
established by Monod, is the so-called Michaelis-Menten form
\BE
f(N)=\mu {N \over N_e +N}
\label{ME}
\EE
which may be also used for the grazing function ($g(P)=\alpha P /
(P_e+P)$). This form may be justified by arguments taken from
enzyme kinetics \cite{murray}. Any function that, like (\ref{ME}),
presents a linear increase close to the origin followed by
saturation at large $N$ receives the name of Hollings type-II
response function. Another commonly used Hollings type-II response
function is the Ivlev one: $g(P) =
\alpha (1-e^{\lambda P})$. Hollings type-III response functions
differ from the above by its behavior at small argument: the
linear increase is replaced by a slower one, indicating that
predators will not be interested in a too small prey
concentration. An example of this type is
\BE
g(P)=\alpha {P^2 \over P_e^2 +P^2} \ \ .
\label{HtIII}
\EE

Large uncertainties exist in the form of the response functions
and in the numerical values of the parameters involved in
(\ref{NPZ}) for particular ecosystems. We mention however that
quantitative determination can be achieved under laboratory
conditions \cite{Hopf}.

Many variations and simplifications of model (\ref{NPZ}) can be
found in the literature. It is rather common to reduce the
modeling to the predator-prey competition of zooplankton and
phytoplankton. In this case the limitation imposed on the
phytoplankton growth by the nutrient concentration is included by
replacing the second equation in (\ref{NPZ}) by one containing a
logistic term:
\BE
F_P(P,Z) = \alpha P\left(1- {P\over C} \right) + ...
\label{logistic}
\EE
The parameter $C$ is called the {\sl carrying capacity}, and
represents the maximum amount of phytoplankton the fluid parcel
can support.

\subsection{Some proposed mechanisms for patchiness}
\label{subsec:patch}

Probably, among the first theoretical approaches used to address
the formation of plankton patches is the one contained in papers
by Skellam\cite{kiSs}, and by Kierstead and Slobodkin\cite{KIsS}
(the so-called KISS theory). They considered the free growth (at
rate $\mu$) of phytoplankton (with concentration $P(\bx,t)$), in a
finite region of water {\sl suitable for growth}, with plankton
escaping the suitable zone by diffusion:
\BE
{\partial P \over \partial t} = \mu P + D \nabla^2 P \ \ .
\label{kissEq}
\EE
This is a very particular case of (\ref{ard}). It is assumed that
$P=0$ outside the region. If the patch is too small, the growth
can not overcome the diffusive escape flux, and the patch dies.
This can be seen, for example, via the exact one-dimensional
solution of (\ref{kissEq}), with boundary conditions such that
phytoplankton disappears outside the region $[0,L]$ of suitable
conditions for growth (i.e. $P(x=0,t)=P(x=L,t)=0$):
\BE
P(x,t)=\sum_{n=1}^\infty A_n \sin\left( {n\pi x \over L} \right)
e^{t\left(\mu - D({n\pi \over L})^2\right)} \ \ .
\label{kissSol}
\EE
The constants $A_n$ are fixed by the initial conditions. It is
clear that if $L<L_c=\pi\sqrt{D/\mu}$ no growth occurs and the
patch dies, so that $L_c$ can be identified with the minimum
patch-size able to support growth.

There is a number of criticisms than can be posed to this
approach. The first one is that the model is incomplete, since
once growth begins it continues without limit. Nonlinear
saturation and interactions with predators would be needed to stop
this. The diffusion coefficient $D$ is certainly not the
originated from the Brownian motion of the organic particles,
since this would be irrelevant to processes above, say, the
millimeter scale. It is rather a turbulent eddy-diffusion
coefficient aimed to represent in an averaged way the effect of
dispersion by the turbulent flow, with values that would depend on
the observation scale as found for example in \cite{okubo}. One
should ask, thus, how can the plankton disperse in such way,
whereas the nutrients remain static within the fixed parcel,
despite they live in the same flow. In any case, if one introduces
a growth rate for phytoplankton of the order of one day ($\mu
\approx 10^{-5}$ s$^{-1}$) and an eddy diffusivity appropriate for the
10-100 km range ($D\approx 100$ m$^2$/s) \cite{okubo}, one finds
$L_c\approx 10$ km, which seemed to compare well with the earliest
observations of plankton patches.

An interesting development that has some common ideas with the
KISS approach, but introduces a new and important element, is
described in \cite{martin}. The idea is that what will control
dispersion of a patch is not only eddy diffusivity, but also the
geometric characteristics of the mean flow. It turns out that if
an incompressible fluid flow induces dispersion in one direction
it necessarily produces concentration in another, to conserve the
fluid volume. The situation is especially clear in a purely
two-dimensional flow, where the rate of expansion of a small fluid
blob in one direction should be exactly the same as the rate of
contraction in another. Thus, an initially circular patch will
take the form of a filament locally aligned with particular
directions in the flow, the unstable manifolds of the flow
hyperbolic points. These manifolds have been nicely observed in
laboratory experiments \cite{Gollub}, and phytoplankton filaments
on the sea surface have been observed from space
\cite{NatureFilament}. Reference \cite{martin} proposes the
following model to describe the transverse profile of a
phytoplankton filament:
\BE
{\partial P \over \partial t} -\lambda x {\partial P \over
\partial x}=
\mu(t) P + D {\partial^2 P \over \partial x^2}  \ \ .
\label{martinEq}
\EE
There are two differences with the KISS model. One is the possible
time-dependence of the growth rate $\mu(t)$, that is a simplified
linear way of modeling nonlinear interactions and predation on
phytoplankton after the first growth stages. The other is the
advective term $-\lambda x
\partial_x P$, associated with a velocity field $v_x(x)=-\lambda x$
that models a local strain that, in the direction perpendicular to
the filament, compresses it. A solution to Eq. (\ref{martinEq})
that is stable and attractive can be found analytically
\cite{martin}: a Gaussian with time-dependent height but fixed
width $w=\sqrt{D/\lambda}$. Thus, the lateral scale of the
filament is not controlled by some externally imposed size of the
suitable water, but by the competition between diffusion and
advection. It is somehow surprising that the biological growth
rate does not affect the filament width in this model. As we will
see in Section \ref{sec:excitable} this feature is not shared by
other models.

A different mechanism that has been invoked as possible source of
plankton patchiness \cite{turing} is the Turing mechanism. It
appears generically when there is competition between a
self-reproducing prey and a predator that diffuses faster than the
prey, and leads to patterns with a characteristic
periodicity\cite{murray}. Although diffusion coefficients of the
organic particles have roughly the same very small value for both
predator (zooplankton) and prey (phytoplankton), the former has
usually a larger vertical mobility. If there is some vertical
shear in the water column, the organisms with larger vertical
mobility will experience a more variable velocity field, and thus
they will be subjected to larger dispersion. If one models such
enhanced dispersion by an effective diffusion coefficient, it
would be larger for zooplankton than for phytoplankton, and then
the Turing mechanism can be at work. The analysis in
\cite{Brindley97} implies that for models of the type (\ref{NPZ}),
the Turing mechanism does not occur in the relevant range of
parameters. But this may not be the case for modifications of this
class \cite{fede}. In addition, related mechanisms, that would
lead also to plankton distributions spatially periodic, may appear
in the presence of differential motion between predator and prey
\cite{rovinsky}.

Horizontal stirring is another mechanism that produces strong
inhomogeneity\cite{abraham}. This is the one that will be
addressed in the following. According to \cite{Brindley94b}, the
attractors of models of the form (\ref{NPZ}) are, for most of the
relevant parameter range, fixed points representing stable species
coexistence. In most cases, at least if type III Hollings
functions are used, excitable behavior \cite{murray} occurs close
to these equilibrium points. This motivates us to consider the
effect of stirring on population models displaying either simple
relaxation to fixed points (Section \ref{sec:dps}) or excitable
behavior (Section \ref{sec:excitable}). Population oscillations
also occur at realistic parameter values \cite{Brindley94b,Hopf},
but we will not consider this situation here. Before going to the
results for the chemical/biological dynamics, we summarize in the
next section some results about {\sl chaotic advection},  which is
the kind of fluid flow considered in the developments presented
here.

\section{The paradigm of chaotic advection}
\label{sec:chaoadvect}
   When dealing with transport processes in complex fluid flows,
the concept of turbulence, the unsteady and irregular kind of flow
in which motion occurs in a large range of scales, is the paradigm
that most often comes to mind. Fluid elements, and transported
particles, follow intricate trajectories in such velocity fields.
It was recognized some time ago \cite{aref} that a turbulent
velocity field is not necessary to produce chaotic fluid-element
trajectories. In three spatial dimensions the dynamical system
(\ref{trajectory}) has chaotic trajectories \cite{Bohr,Ott} even
for very simple steady and smooth velocity fields $\bv(\bx)$. In
two-dimensional flows, a simple periodic time dependence in
$\bv(\bx,t)$ is enough to induce generically chaotic trajectories
in (\ref{trajectory}). This situation of chaotic trajectories
produced by a simple and regular velocity field is called {\sl
chaotic advection} or {\sl Lagrangian turbulence}. The essential
characteristic of chaos is sensibility to initial conditions. This
means that fluid elements initially very close would follow
diverging trajectories. The exponential rate of separation is
given by the flow Lyapunov exponent $\lambda$. In an
incompressible fluid, contraction must occur in another direction
to conserve volume. The clearest situation is in hyperbolic
two-dimensional flows, where one can define at each point an
expanding direction and a contracting direction, with both the
contraction and expansion rate occurring at the same exponential
rate $\lambda$. As a consequence, portions of fluid initially
compact become stretched in long and thin filaments, and are also
repeatedly folded to remain inside the system. As a result,
chaotic advection produces a ``cascade'' of inhomogeneities from
large scales to smaller and smaller scales, in the same
qualitative way as a fully turbulent velocity field will do.
Quantitative details are, however, different\cite{Falko}. It turns
out that the patterns produced by advection in smooth velocity
fields are more singular than the ones produced by singular
velocity fields such as Kolmogorov turbulence.

Experimental realizations of (nonturbulent) chaotic advection can
be performed in small containers with high viscosity fluids, so
that Reynolds number remains small and the flow laminar. It is now
recognized that the regime of chaotic advection also applies to
turbulent flow at scales smaller than the Kolmogorov scale ( $l_K
\equiv (\nu^3/\epsilon)^{1/4}$, where $\nu$ is the kinematic viscosity and
$\epsilon$ the energy dissipation rate). This is the so-called
{\sl Batchelor regime}\cite{batchelor}. The Lyapunov exponent in
that range is of the order of the mean strain $\lambda\approx
\sqrt{\epsilon/\nu}$. For water ($\nu
\approx 10^{-6}$ m$^2$s$^{-1}$) at conditions typical in the upper
mixed layers of oceans and lakes ($\epsilon \approx 10^{-7}$ W
kg$^{-1}$,\cite{dissipation}), $l_K$ is between the millimeter and
the centimeter. Thus the Batchelor regime is certainly relevant
for plankton processes such as feeding, aggregation, encounter,
etc. It is however a range much smaller than the structures shown
in Fig.~\ref{fig:transect} or the ones visible from satellites. A
third situation where chaotic advection has been used with success
is in the modeling of the largest scales of geophysical flows
\cite{peter}. Smaller motions would then have to be represented by
some eddy diffusion.

In the following, we focus on the situation of chaotic advection.
This would allow us to progress in our objective of identifying
mechanisms that can contribute to classify observations in robust
scenarios. Chaotic advection is a framework simple enough to
obtain a number of explicit results that turn out to be rather
insensible to model details. Of course, explanation of particular
geophysical observations would require additional discussion on
whether or not the paradigm of chaotic advection applies to the
scales considered.

\section{The decaying transported substance: a model for stable dynamics}
\label{sec:dps}

Most of the simpler models of chemical dynamics, and models such
as (\ref{NPZ}) in a large range of parameters, evolve in such a
way that, at long times, the different species reach an
equilibrium at which they coexist at some fixed-point
concentration values. When some of the coefficients in
Eq.~(\ref{NPZ}) are functions of space they act (when coupled with
the transport equation (\ref{trajectory})) as time-dependent
source or sink forcing terms that disturb such equilibrium. Even
in this forced case it may happen that the concentration values at
each fluid element tend to relax to a unique time-history,
determined by the fluid-element trajectory. When this happens we
say that we have a {\sl stable chemical dynamics}. In a sense, the
concentrations try to relax to the local-equilibrium values that
would correspond to the values of the parameters found along the
fluid-element trajectory. The mathematical way to check this
stability is by initializing a particular fluid element with two
slightly different sets of concentration values, and monitor how
these values evolve under the coupled chemical and transport
dynamics (\ref{NPZ}) and (\ref{trajectory}). The exponential rate
of divergence between these two different initial concentrations,
under the same fluid trajectory, defines the so-called {\sl
chemical Lyapunov exponent}, $\lambda_C$. If this quantity is
negative, it identifies {\sl convergence} to a common evolution,
and this is the case of {\sl stable chemical dynamics}. The usual
ergodic properties would guarantee that the same value of the
chemical Lyapunov exponent will be obtained for almost all fluid
trajectories and initial concentrations.

The simplest case of {\sl stable chemical dynamics} is the {\sl
linearly decaying passive scalar}, or {\sl first order reaction}.
It consists in the evolution of a single concentration $C(\bx,t)$
under a linear decay at rate $b$:
\begin{equation}
\frac{\partial C}{\partial t}+{\bf v}({\bx},t) \cdot \nabla C=
S({\bx})- b C + D \nabla^2 C,
\label{Euler}
\end{equation}
A space-dependent forcing consisting in a source $S(\bx)$ of the
substance is included, to maintain a non-trivial concentration
field at long times.

It is easy to check (by using the Lagrangian representation valid
for $D\rightarrow 0$) that $\lambda_C=-b$. The simple dynamics
(\ref{Euler}) can be considered either as an approximation to more
complex chemical or biological evolutions with stable chemical
dynamics, or as a description of simple specific processes such as
spontaneous decomposition of unstable radicals, evolution of a
radioactively or photochemically decaying substance, or relaxation
of sea-surface temperature towards atmospheric values
\cite{Abraham2}. It also describes the concentration of the $C$
reactive in the binary reaction $B+C\rightarrow D$ when the $B$
concentration is maintained constant.

The first systematic study of model (\ref{Euler}) seems to be in
\cite{corrsin}. The author analyzed, by a spectral element method,
the power spectrum in a number of threedimensional turbulent
regimes. In the Batchelor regime a power-law is found with an
exponent depending on the quotient between the decay rate $b$ and
the mean strain rate, which is equivalent to a flow Lyapunov
exponent $\lambda$. Here we will obtain an essentially equivalent
result within the framework of chaotic advection\cite{PRL,PRE}. In
addition, intermittency corrections and the differences between
open and closed flows will be discussed. Our results are first
obtained in terms of the behavior of the concentration differences
between spatially close points ($\delta C=C(\bx+\bdr,t)-C(\bx,t)$)
and later in terms of the structure functions, that are the
moments of $\delta C$ averaged in space.

Exact results in \cite{Chertkov98} for the decaying scalar in a
closed random Kraichnan flow in the Batchelor regime confirm the
generality of the mechanism found: chaotic advection and decaying
chemistry produce singular patterns that can be characterized by
scaling exponents depending on the chemical decay rate and on the
flow Lyapunov exponent (more precisely, on the finite-time
Lyapunov exponent distribution). We will present here also results
for nonlinear plankton models, showing that the results of the
linearly decaying chemistry can be translated to the case of
nonlinear stable chemistry {\sl at the smaller scales} simply by
replacing the decay rate $b$ by the absolute value of the chemical
Lyapunov exponent $|\lambda_C|$. Detailed justification of this
can be found in \cite{CHAOS} for multiple species in the framework
of chaotic advection, or in \cite{Chertkov99} for a single
concentration in the framework of the random Kraichnan flow. It is
important to mention that in the present framework the relevant
time scale affecting the distributions structure is identified to
be $\lambda_C$, and not other time scales frequently used in the
literature such as the phytoplankton linear growth rate.

\subsection{The smooth-filamental transition for the decaying passive scalar
in a closed flow}
\label{subsec:transition}

We consider (\ref{Euler}) with a two-dimensional, incompressible,
smooth, and non-turbulent velocity field. Chaotic advection is
obtained generically if a simple time-dependence is included in
${\bf v}({\bx},t)$. We assume that diffusion is weak and transport
is dominated by advection. Thus one expects that the distribution
on scales larger than a certain diffusive scale is not affected by
diffusion, and set $D=0$. In this limit the above problem can be
described in a Lagrangian picture by the pair of dynamical systems
(\ref{trajectory}) and
\begin{equation}
\frac{d C}{dt}=S[\bx=\hr(t)]-b C,\;
\label{LagrangeC}
\end{equation}
where the solution of (\ref{trajectory}) gives the trajectory of a
fluid parcel, $\hr(t)$, while (\ref{LagrangeC}) describes the
Lagrangian chemical dynamics in this fluid element: $ C(t)
\equiv C({\bx}=\hr(t),t)$.

To obtain the value of the chemical field at a selected point
$\bx$ at time $\bar t$ one needs to know the previous history of
this fluid element, that is the trajectory $\hr(t)$ ($0 \le t \le
\bar t$) with the final condition $\hr(\bar t) = \bx$. This can be
obtained by the integration of (\ref{trajectory}) backwards in
time. Once $\hr(t)$ has been obtained, the solution of
(\ref{LagrangeC}) is
\begin{equation}
C(\bx,\bar t) = C[{\br}(0),0] e^{-b \bar t} + \int_0^{\bar t}
S[{\br}(t)] e^{-b (\bar t-t)} dt.
\label{field}
\end{equation}

Then one can obtain the difference at time $\bar t$ of the values
of the chemical field at two different points $\bx$ and $\bx+\bdr$
separated by a small distance $\bdr$ in terms of the difference
$\delta C[\hr(t),t;\hdr(t)] \equiv C[\hr(t) +
\hdr(t),t]-C[\hr(t),t]$ for $0 \le t\le \bar t$, namely:
\BE
\delta C(\br,\bar t; \bdr) = \delta C[\hr(0),0;\hdr(0)] e^{-b \bar
t} + \int_0^{\bar t} \delta S[\hr (t); \hdr(t)] e^{-b (\bar t- t)}
dt
\label{deltaC1}
\EE
where $ {\delta {\br}}(t)$ ($0 \le t \le \bar t$) is the
time-dependent distance between the two trajectories that end at
$\bx$ and $\bx+\bdr$ at time $\bar t$, and $\delta S$ is the
difference of the source term at points $\hr(t)$ and
$\hr(t)+\hdr(t)$. In the following we omit the first term in the
right-hand-side of (\ref{deltaC1}) since it always disappears at
long times.

If the trajectory is chaotic, we have in the backwards dynamics,
for $t<0$ and large,
 $|\hdr (t)| \sim \left|
\hdr(0) \right| e^{-\lambda t}$, where
$\lambda$ is the positive value of the Lyapunov exponent along the
trajectory. In (\ref{deltaC1}), $\hdr$ is obtained backwards
starting from $\bdr$ at $t=\bar t$. In this case:
\begin{equation}
\hdr(t) \approx  \bdr \ e^{\lambda(\bar t
-t)}\ , t<\bar t\ .
\label{dispersion}
\end{equation}
We note that there is a particular direction for the orientation
of the initial displacement $\bdx$ along which the trajectories
converge instead of diverging (this is the contracting direction
in the backwards flow which is the expanding one in the forward
dynamics). For $\bdx$ oriented along it, $-\lambda$ in
(\ref{dispersion}) should be replaced by $\lambda$. For hyperbolic
dynamical systems, the value of the Lyapunov exponent is the same
for {\sl almost all} the initial or final conditions $\bx$
\cite{Bohr,Ott}. We will show later, however, that the deviations
that could occur in sets of zero measure may have some observable
consequences.

Taking the limit
 ${\bdx} \rightarrow 0$ and with
substitution of (\ref{dispersion})  in  Eq.~(\ref{deltaC1}), one
obtains
\BE
\delta C(\br,\bar t; \bdr) \approx   \bdr \cdot  \int_{0}^{\bar t}
 \nabla S [\hr(t)] e^{(\lambda -b)(\bar t-t)}
dt \ \ .
%%\right]
\label{deltaCfinal}
\EE
 Rewriting $\bdr= \bar {\bf n}
|\bdr|$ so that $\bar {\bf n}$ is a unit vector, one finds the
directional derivative along the direction of $\bar {\bf n}$ as
\BE
\bar {\bf n} \cdot \nabla C(\br,\bar t)\approx \bar {\bf n} \cdot
 \int_{0}^{\bar t} \nabla S
[\hr(t)] e^{(\lambda -b)(\bar t-t)} dt  \ .
\label{directional}
\EE
If $\lambda < b$ this derivative remains finite in the $\bar t
\rightarrow \infty$ limit and the asymptotic field $C_\infty(\bx)
\equiv C({\bx},\bar t \rightarrow \infty)$ is smooth
(differentiable). Otherwise the derivatives of $C$ diverge as
$\sim e^{(\lambda -b)\bar t}$ leading to a nowhere-differentiable
irregular asymptotic field. The exception occurs when $\bdr$
points along the expanding direction of the forward flow:
$\lambda$ should be replaced by $-\lambda$ and the directional
derivative is always finite. The irregular field, thus, has a
filamental character. It should be noted that the limiting
distribution $C_\infty$ is not a steady field, but one following
the time dependence of the flow. For time-periodic flows ${\bf
v}({\bx},t)$, $C_\infty$ will also be time periodic. Its singular
characteristics however do not change in time.

The characterization of the singular asymptotic field has been
performed in \cite{PRE}. Defining $\delta C_\infty \equiv \delta
C({\bx},\bar t \rightarrow
\infty)$ and taking into account that the exponential separation (\ref{dispersion})
can only persist until it reaches the characteristic scale of the
flow field, one finds \cite{PRE} in the $|{\bdx}| \rightarrow 0$
limit the scaling
\begin{equation}
\delta C_\infty (\br;\bdr) \sim |\bdr|^{\alpha} \ ,
\label{Holder1}
\end{equation}
with a H\"older exponent $\alpha$ given by
\begin{equation}
\alpha = \min \left\{ \frac{b}{\lambda} , 1 \right\}.
\label{Holder2}
\end{equation}
This general relation expresses the local H\"older exponent in
terms of the local infinite-time Lyapunov exponent and the
chemical decay rate. As stated before, for hyperbolic systems this
Lyapunov exponent has the value $\lambda_0$ everywhere but in a
set of zero measure. By decreasing $b$ or increasing $\lambda$ one
finds a transition from a smooth $\alpha=1$ distribution to a
rough $\alpha<1$ one with filamental structure. This is the
so-called {\it smooth-filamental transition}\cite{PRL}. For
uniform $\lambda=\lambda_0$, the relationship between the
H\"{o}lder exponent and the power spectrum decay exponent ($\beta$
in the power spectrum scaling $S(k) \sim k^{-\beta}$) is
$\beta=1+2\alpha$. Thus the result (\ref{Holder2}) implies an
exponent $\beta$ larger than one, as it is usually observed. The
Batchelor result \cite{batchelor} $S(k) \sim k^{-1}$, valid for
the nondecaying scalar in smooth velocity fields is recovered for
$b=0$.

\begin{center}
\begin{figure}[t]
\epsfig{file=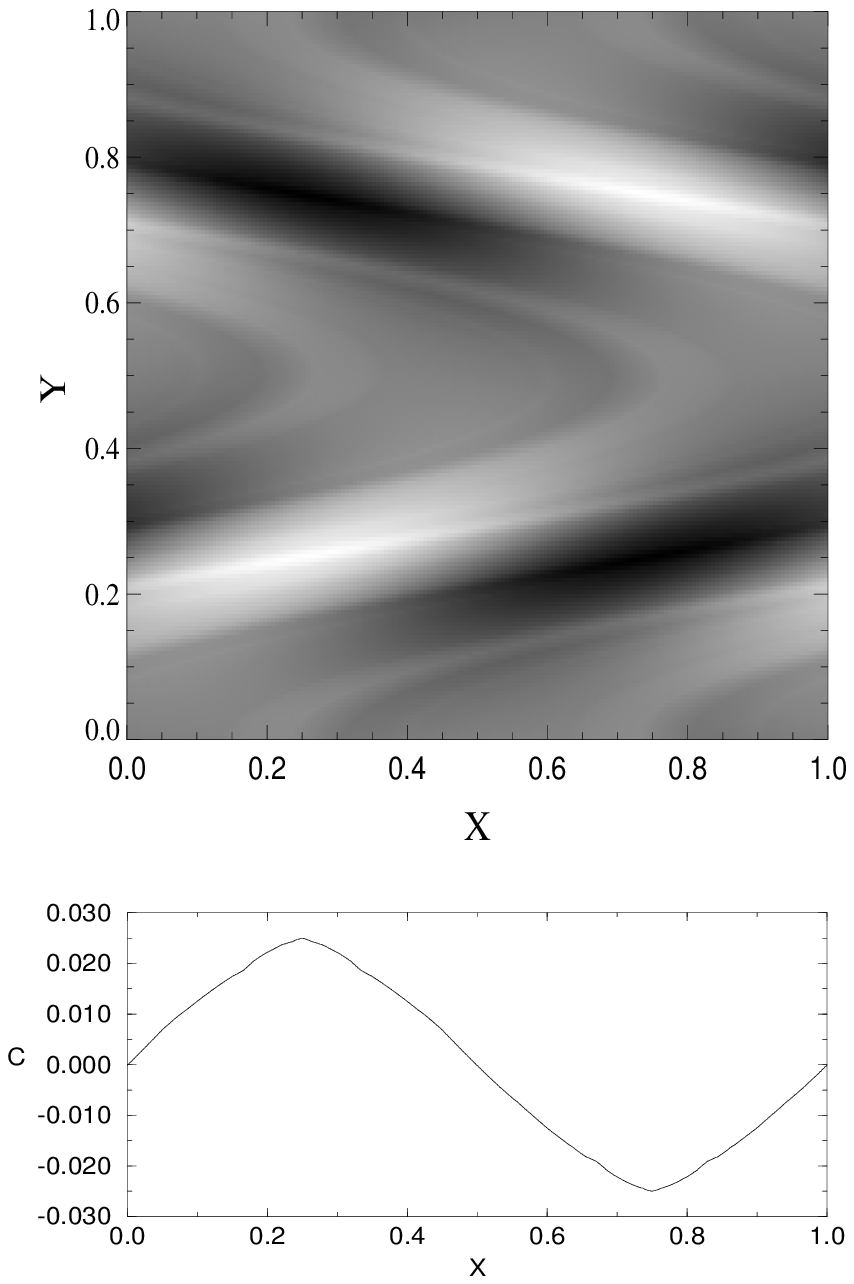,width=.45\linewidth}
\epsfig{file=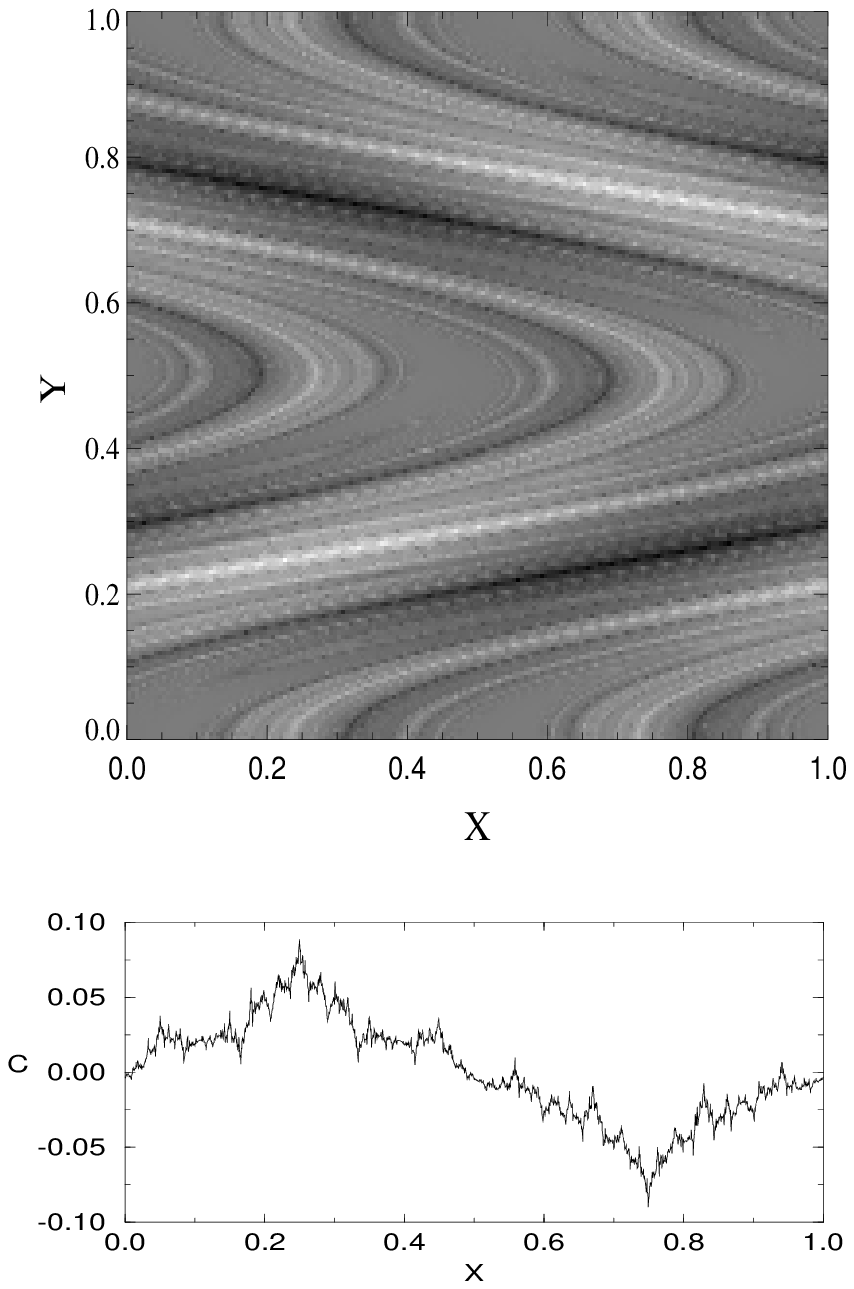,width=.45\linewidth}
%\vskip 0.5cm
\caption{\protect\footnotesize Snapshots of chemical patterns of the
forced decaying scalar under the flow (\ref{flow}). Darker grey
levels indicate smaller concentrations. The lower panels are
horizontal cuts along the line $y=0.25$. Left: $b=4.0$ and
$T=1.0$, so that $\lambda_0=2.67<b$. A smooth pattern is seen, in
agreement with the theoretical arguments. Right: $b=0.1$ and
$T=1.0$ so that $\lambda_0>b$ and a filamental pattern is
obtained. The horizontal cut clearly displays the fractal nature
of the field.}
\label{fig:closed}
\end{figure}
\end{center}

In Fig.~\ref{fig:closed} we present snapshots of the asymptotic
field $C_\infty$ evolving according to (\ref{trajectory}) and
(\ref{LagrangeC}). For the flow we take a simple time-periodic
velocity field defined in the unit square with periodic boundary
conditions by
\begin{eqnarray}
v_x(x,y,t)& =& -\frac{2U}{T} \Theta \Bigl(\frac{T}{2}-t \bmod T
\Bigr) \cos({2\pi y}) \nonumber \\ v_y(x,y,t)&=& -\frac{2U}{T}
\Theta \Bigl(t \bmod T-\frac{T}{2} \Bigr) \cos({2\pi x})
\label{flow}
\end{eqnarray}
where $\Theta(x)$ is the Heavyside step function. In our
simulations $U=1.2$, which produces a flow with a single connected
chaotic region  in the advection dynamics. The value of the
numerically obtained Lyapunov exponent is $\lambda_0 \approx
2.67/T$. We use the source term $S(x,y)=1+\sin{(2 \pi x)}\sin{(2
\pi y)}$. Backward trajectories with initial coordinates on a
square grid were calculated and used to obtain the chemical field
at each point by integrating (\ref{LagrangeC}) forward in time.
The smooth and the filamental behavior are found by changing the
decay rate $b$, in complete agreement with the theory.

\subsection{Smooth-filamental transition in a nonlinear plankton
dynamics model}
\label{subsec:plankton}

As we mentioned before, all the results presented in the previous
section are valid for a more complex but stable chemical or
biological dynamics, as far as the decay rate $b$ is replaced the
chemical Lyapunov exponent $|\lambda_C|$. In this subsection we
illustrate the smooth-filamental transition with a simple model of
plankton dynamics immersed in a meandering jet flow\cite{PCE}.
Additional results for nonlinear models can be found in
\cite{CHAOS}.

As in the previous subsection, diffusion is neglected so that a
Lagrangian representation in terms of a {\sl chemical} and a {\sl
transport} dynamical system is possible. The {\sl chemical}
plankton model is the one used in \cite{abraham}. It is similar to
(\ref{NPZ}), but with phytoplankton growth ruled by a logistic
term (\ref{logistic}), and a relaxational dynamics imposed on the
carrying capacity $C$:
\BA
\frac{dC}{dt} &=& \gamma \left(   C_0({\bx})-C \right)
\nonumber \\
\frac{dP}{dt} &=& P\left( 1-\frac{P}{C} \right)-PZ
\nonumber \\
\frac{dZ}{dt} &=& PZ-\delta Z^2
\label{CtPtZt}\ .
\EA
In the absence of flow, the only stable fixed point of model
(\ref{CtPtZt}) is given by $C^*=C_0({\bx})$, $P^*= {C_0
\delta }/({\delta + C_0})$, and $Z^*={P^*}/{\delta}$.

The {\sl transport} dynamical system is
\BA
\frac{dx}{dt} &=&-\frac{\partial\psi}{\partial y}  \nonumber \\
\frac{dy}{dt} &=& \frac{\partial\psi}{\partial x} \ \ ,
\label{streamflow}
\EA
given in terms of the following streamfunction \cite{bower}:
\BE
\psi(x,y)   =   1 - \tanh \left(\frac{y-B(t)
\cos\left[ k(x-ct)\right]}{\left( 1+k^2 B(t)^2 \sin^2
\left[k(x-ct)\right] \right)^{\frac{1}{2}}} \right).
\label{jetstreamfunction}
\EE
It describes a jet flowing eastwards, with meanders in the
North-South direction. These meanders are also advected by the jet
at a phase velocity $c$. $B(t)$ and $k$ are the (properly
adimensionalized) amplitude and wavenumber of the undulation in
the streamfunction. As we are interested in a closed flow, we
impose periodic boundary conditions at the ends of the interval
$-L_x<x<L_x$. Particles leaving the region through the right
boundary are reinjected from the left. Nutrients are injected in
and out continuously from fluid elements as these traverse the
different regions of the carrying capacity source
\begin{equation}
C_0(x,y)=1+A \sin(2\pi x/L_x)\sin(2\pi y/L_y).
\label{Csource}
\end{equation}

Chaotic advection appears in this model if $B$ is made to vary in
time, for example periodically: $B(t)=B_0+\epsilon \cos(\omega
t+\theta)$. In our calculations we use  the parameter values
$B_0=1.2$, $c=0.12$, $k={2\pi}/{L_x}$, $L_x=7.5$, $L_y=4.0$,
$\omega=0.4$, $\epsilon=0.3$, and $\theta =\frac{\pi}{2}$. These
values guarantee the existence of `large scale chaos', i.e, the
possibility that a test particle crosses the jet passing from
North to South or viceversa. This is weaker than the requirement
of hyperbolicity, but is enough to illustrate the general aspects
of our theory. In the chemical subsystem we use $A=0.2$, and
$\delta=2.0$, and vary the value of $\gamma$, which controls the
relaxational time-scale. It turns out that the resulting chemical
dynamics is stable \cite{PCE}.

\begin{center}
\begin{figure}[t]
\epsfig{file=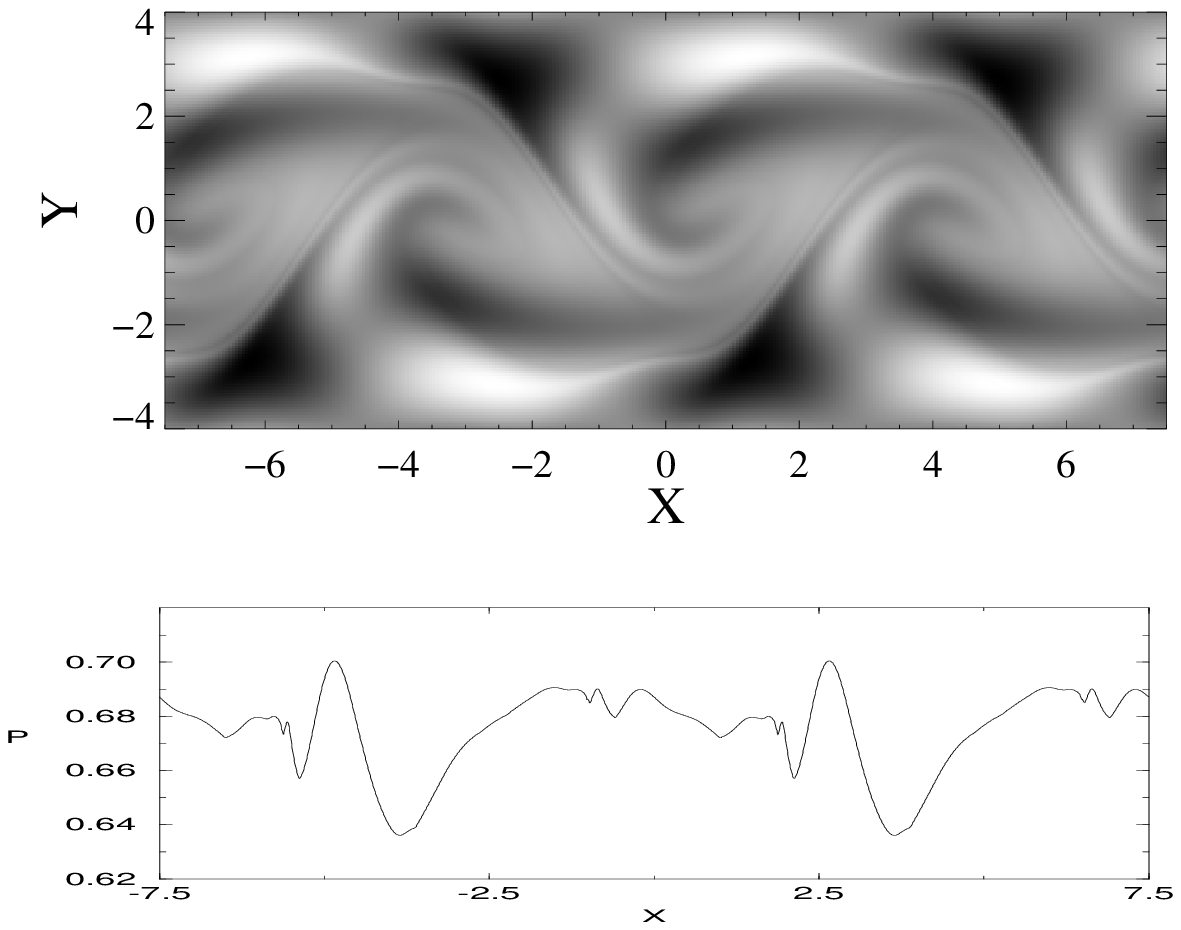,width=.45\linewidth}
\epsfig{file=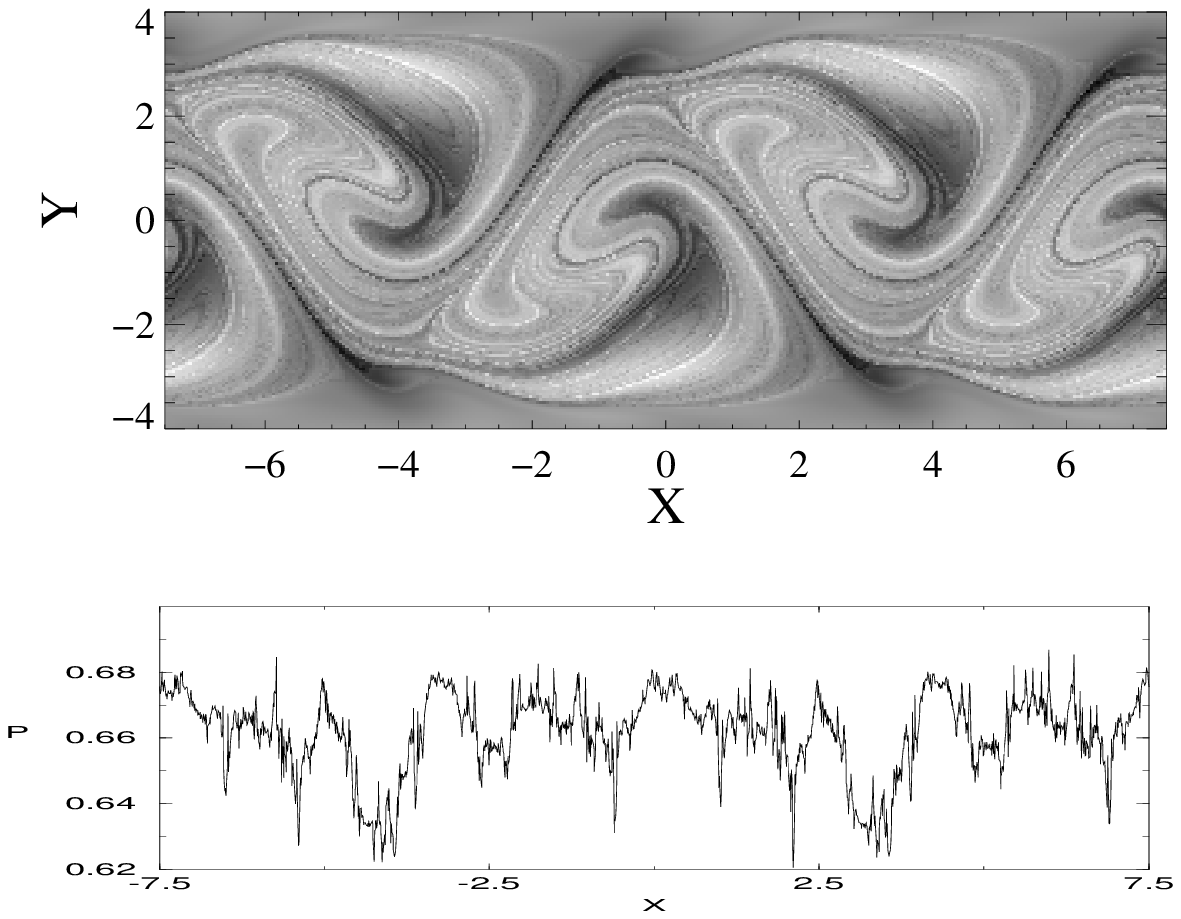,width=.45\linewidth}
\vskip 0.5cm
\nopagebreak
\caption{\protect\footnotesize Snapshots of phytoplankton patterns from
(\ref{CtPtZt}) and (\ref{streamflow}). The lower panels are
transects along $y=0.8$. Left: $\gamma=0.25$; a smooth pattern is
obtained. Right: for smaller relaxation rate $\gamma=0.025$, a
fractal filamental pattern is generated.}
\label{figure12}
\end{figure}
\end{center}

Fig.~\ref{figure12} shows snapshots of the long-time phytoplankton
distributions. The smooth-filamental transition is clearly
observed. More quantitative discussions can be found in reference
\cite{CHAOS}.

\subsection{Open flows}
\label{subsec:open}

A finite region is said to be traversed by an open flow if almost
all fluid elements enter the region and leave it forever in a
finite time. A prototype is a stream passing around a cylindrical
body. If the inflow velocity is high enough vortices form in the
wake of the cylinder and make the flow time-dependent in this
region, while the flow remains steady in front of the cylinder or
in the far downstream region. Advected particles enter the
unsteady region, undergo transient chaotic motion
\cite{Tel,Jung,Ziemniak,Sommerer}, and finally escape and move
away downstream on simple orbits. The time spent in the mixing
region, however, depends strongly on the initial coordinates, with
singularities on a fractal set corresponding to particles trapped
forever in the mixing region. This is due to the existence of a
non-attracting chaotic saddle (a fractal object of zero measure)
formed by an infinite number of bounded hyperbolic orbits in the
mixing region. The stable manifold of this chaotic saddle contains
orbits coming from the inflow region but never escaping from the
mixing zone. Thus, permanent chaotic advection is restricted to
this fractal set of zero Lebesgue measure.
 Points close to the unstable manifold of the chaotic
saddle have spent a long time in the mixing region of the flow
moving near chaotic orbits with a positive Lyapunov exponent. For
points precisely on this unstable manifold, the backwards
trajectories (the ones from which the Lyapunov exponent in
(\ref{Holder2}) should be computed) remain in the chaotic saddle,
thus leading to $\lambda_0>0$. The other trajectories spend in the
mixing region just a finite time (both in the forward as in the
backwards time direction), so that they can not contribute to the
development of singularities in chemical distributions
((\ref{directional}) is singular only in the $\bar t \rightarrow
\infty$ limit). In fact almost all trajectories are characterized
by a long-time Lyapunov exponent equal to zero.

Thus open flows provide a rather clear example of strong
space-dependence
%%, or intermittency,
of Lyapunov exponents. According to Eq.~(\ref{Holder2}), the
H\"{o}lder exponent may be different from $1$ only on the unstable
manifold of the chaotic saddle, thus implying that the transition
from smooth to filamental structure now only takes place in this
fractal set of zero measure. The background chemical field is
always smooth, independently of the value of $b$.

\begin{figure}[t]
\epsfig{file=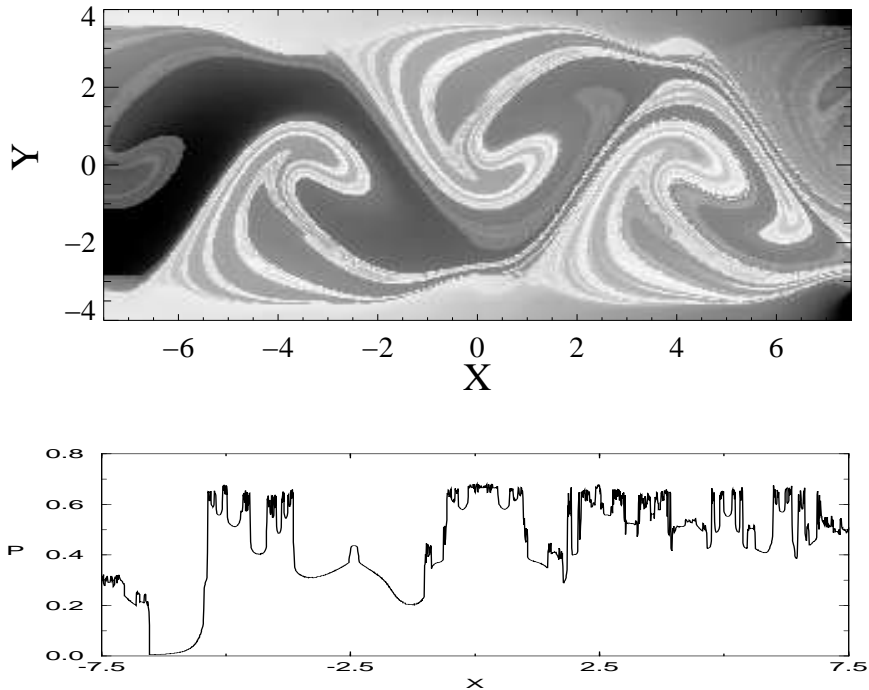,width=.8\linewidth}
\vskip -1cm
\nopagebreak
\caption{\protect\footnotesize Phytoplankton filamental pattern in the open
jet, with a transect along $y=0.8$. }
\label{figure3}
\end{figure}

Patterns of chemically decaying substances in a cylinder wake are
presented in \cite{PRE}. Here we present the case of the stable
biological model (\ref{CtPtZt}) under the jet flow
(\ref{streamflow})-(\ref{jetstreamfunction}). It is made open
simply by not imposing the periodic boundary conditions of the
previous subsection. The chaotic mixing region is not restricted
to a finite region, but we restrict the source forcing
(\ref{Csource}) to $-L_x<x<L_x$. This is the region shown in
Fig.~\ref{figure3}, being $C_0=0$ outside. We let the fluid
particles to enter this region with very small $C$, $P$ and $Z$
concentrations. Most of them remain in the forcing region only for
a finite time, so that, as in the cylinder wake case, they do not
develop singularities. Only particles remaining in the chaotic
saddle, i.e. the set of nonscaping orbits, forever will lead to
diverging gradients. Fig.~\ref{figure3} shows a phytoplankton
pattern for the same parameter values as before and $\alpha =
0.025$. Smooth zones coexist with singular features. The behavior
is distinct from the closed flow case (Fig.~\ref{figure12}) and
represents a new scenario for the development of chemical or
biological patterns.

Since the irregularities now appear only on a set of measure zero,
one could ask if they can have any significant effect on
measurable quantities. In order to clarify this, we consider for
simplicity the case of a single chemical reactive $C$, and instead
of the previous characterization of the point-wise strength of the
singularities by the H\"older exponent, let us investigate the
scaling with distance $\delta x \equiv |\delta{\bx}|$ of the
spatial average of the differences $\delta C_\infty$. For
simplicity, let us assume that, on the saddle, there is no
distribution in the local infinite-time Lyapunov exponents, i.e.,
that the advection on the chaotic saddle is characterized by a
single Lyapunov exponent $\lambda_0$. In this case the partial
fractal dimension (i.e. the dimension of intersections of the set
with a line) of the manifolds of the chaotic saddle is
\cite{Bohr,Jung}
\begin{equation}
\tilde D = 1 - {\kappa \over \lambda_0}.
\end{equation}
Here $\kappa$ is the escape rate, that is the rate of the
exponential decay ($\sim e^{-\kappa t}$) of the number of fluid
elements spending time longer than $t$ in the mixing or in the
forcing region. On a one-dimensional transect of unit length the
total number of segments of length $\delta x$ is $(\delta x)^{-1}$
while the number of segments containing parts of the unstable
manifold (with partial fractal dimension $\tilde D$) is  $\sim
{\delta x}^{-\tilde D}$. Thus, according to (\ref{Holder2}) the
spatial average of $\delta C_\infty$ along this line, $\langle
\delta C_\infty({\bf x};\delta x) \rangle$, can be written as
\begin{eqnarray}
\langle \delta C_\infty({\bx};\delta x) \rangle = (\delta x)
(\delta x)^{-\tilde D} (\delta x)^{b/\lambda_0} + \nonumber \\ +
(\delta x) [ (\delta x)^{-1} - (\delta x)^{-\tilde D} ] (\delta x)
\label{preopenaverage}
\end{eqnarray}
where the first term pertains to the singular component
($\alpha=b/\lambda_0$), while the second one pertains to the
smooth component ($\alpha=1$). In the limit $\delta x
\rightarrow 0$ the dominating behavior is
\begin{equation}
\langle \delta C_\infty({\bx};\delta x) \rangle \sim \delta
x^\zeta\ \ ,
\label{openaverage}
\end{equation}
with
\begin{equation}
\zeta=\min \left\{ 1, 1+b/ \lambda_0-\tilde D \right\}  =
\min \left\{1, {{b + \kappa} \over
{\lambda_0}} \right\}
\label{zeta}
\end{equation}
showing that if $\tilde D < {b \over \lambda_0}$ (or, $b+\kappa >
\lambda_0$) the average will be dominated by the smooth component,
but if the fractal dimension of
 the singular set is large enough it contributes to
the scaling of $\langle \delta C_\infty({\bf r};\delta r)
\rangle$. We see that moments of $\delta C$ may be sensible to
fractal inhomogeneity in $\alpha$, or {\sl intermittency}.

\subsection{Anomalous scaling of structure functions}
\label{subsec:structure}

The strongly intermittent structure of singularities in open flows
is an extreme example. There are additional inhomogeneities
affecting both to the open and to the closed flows: although, in
the long-time limit the Lyapunov exponent is the same for almost
all trajectories in an ergodic region, deviations can persist on
fractal sets of measure zero, and as we saw above such sets can
contribute significantly to the global scaling. The origin of
these inhomogeneities can be traced back by analyzing the
finite-time distribution of Lyapunov exponents. In general, the
finite-time stretching rates, or local Lyapunov exponents
\cite{Ott}, have a certain distribution around the most probable
value. This distribution approaches the time-asymptotic form
\cite{Bohr,Ott}:
\begin{equation}
P(\lambda, t) \sim t^{1/2} e^{-G(\lambda)t}
\label{finitelyap}
\end{equation}
where $G(\lambda)$ is a function characteristic to the advection
dynamics, with the property that $G(\lambda_0)=G'(\lambda_0)=0$
and $G(\lambda)>0$, and $\lambda_0$ is the most probable value of
the Lyapunov exponent. At infinitely-long times all the measure
becomes concentrated at this single value $\lambda_0$, as stated
before. There are however fractal sets of zero measure that do not
share this unique value. The partial dimension of the set with
Lyapunov exponent value $\lambda$ is \cite{PRE} $\tilde D(\lambda)
= 1-G(\lambda)/\lambda$. The coexistence of such different and
interwoven fractal sets is the signature of {\sl multifractality}.

For a robust quantitative characterization of the filamental
structures, accessible to measurements and sensible to the
intermittency features, we consider now the scaling properties of
the structure functions associated with the chemical fields. For a
single species $C$, the $q$th order structure function is defined
as
\begin{equation}
S_q(\delta x) = \langle |\delta C_\infty({\bx};\delta x)|^q
\rangle
\label{Sq}
\end{equation}
where $\langle ... \rangle$ represents averaging over different
locations ${\bf x}$, and $q$ is a parameter (we will only consider
structure functions of positive order ($q>0$)). In the absence of
any characteristic length over a certain range of scales the
structure functions are expected to exhibit, as $\delta x
\rightarrow 0$, a power-law dependence
\begin{equation}
S_q(\delta x) \sim \delta x^{\zeta_q}
\label{zetaq}
\end{equation}
characterized by the set of scaling exponents $\zeta_q$. The
scaling exponent of the power spectrum is given by
$\beta=1+\zeta_2$.

If the H\"older exponent of the field has the same value
everywhere, given by (\ref{Holder2}) with $\lambda=\lambda_0$, the
scaling exponents of the resulting {\it mono-affine} field are
simply
\begin{equation}
\zeta_q= q \alpha_0 = q {b \over \lambda_0}.
\end{equation}
(we have assumed $b<\lambda_0$). In general, the fractal sets of
partial dimensions $\tilde D(\lambda)$ should be taken into
account in an average such as (\ref{preopenaverage}) but for a
continuous range of values of $\lambda$. The result for the
scaling exponents in (\ref{zetaq}) is \cite{PRE}:
\begin{equation}
\zeta_q = \min_{\lambda} \left \{q, 1+{q b \over \lambda} -
\tilde D(\lambda) \right \} = \min_{\lambda} \left \{q, {q b+G(\lambda)
\over \lambda} \right \}
\label{scalingexp}
\end{equation}
Equation (\ref{zeta}) is a particular case of (\ref{scalingexp})
for $q=1$ and in the approximation of considering a single value
of $\lambda$ on the chaotic saddle.

\begin{center}
\begin{figure}[t]
\epsfig{file=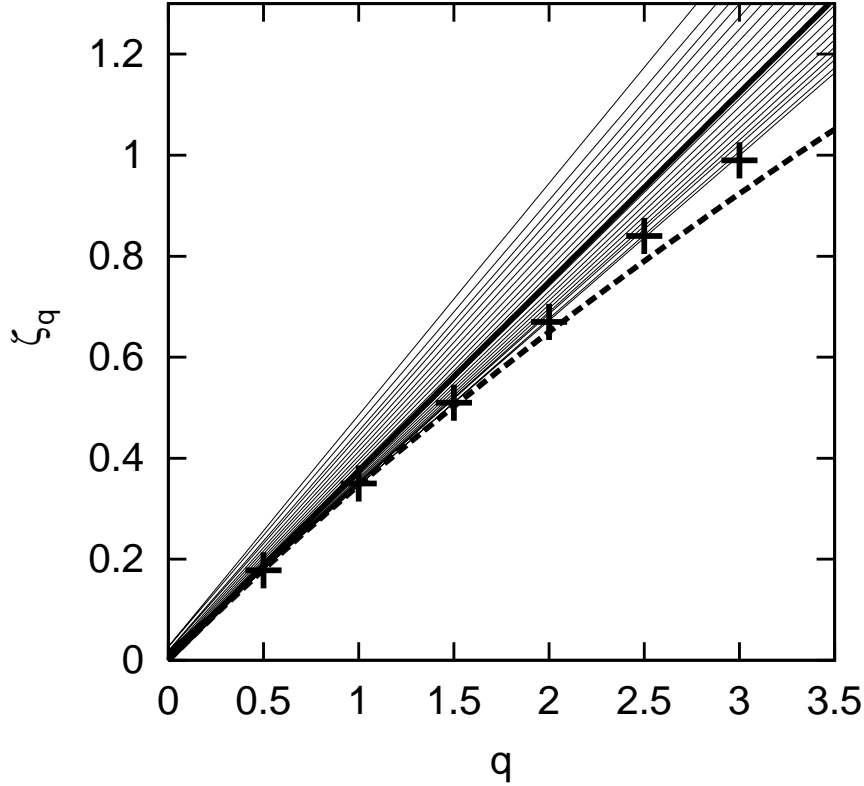,width=.7\linewidth}
\vskip 0.5cm
\nopagebreak
 \caption{\protect\footnotesize The scaling
exponents $\zeta_q$ for the structure functions of the linearly
decaying scalar under the flow (\ref{flow}). Same parameters as in
Fig.~\ref{fig:closed} except $b=1.0$. Thick line: the mono-fractal
approximation $\zeta_q=q b/\lambda_0$. Thin lines: the curves
$\zeta_q=q$ and $\zeta_q=[qb+G(\lambda)]/\lambda$, for different
values of $\lambda$; for the numerical values of $G(\lambda)$ see
\protect\cite{PRE}. According to Eq.~(\protect\ref{scalingexp}), the
actual values of the scaling exponents are given by the lower
envelope of this set of curves. This is confirmed by the
numerically determined values of $\zeta_q$ (crosses). Dashed line:
the approximation (\protect\ref{Chertkovscaling}).  }
\label{fig:scaling}
\end{figure}
\end{center}

Multifractality thus originates anomalous scaling (nonlinear
dependence of $\zeta_q$ on $q$) of the structure functions.
Multifractality and anomalous scaling have been observed in real
data from plankton and temperature distributions (see for example
\cite{Pascual,Seuront,Abraham2}. Here we present the scaling
exponents for structure functions of the linearly decaying scalar
(\ref{LagrangeC}) under the flow (\ref{flow}).
Fig.~\ref{fig:scaling} shows the numerically calculated scaling
exponents (i.e. obtained by direct application of Eq.~(\ref{Sq})),
and the family of lines corresponding to (\ref{scalingexp}) for
different values of $\lambda$. Eq.~(\ref{scalingexp}) predicts
that the actual values of $\zeta_q$ are given by the lower
envelope of the set of lines, a fact confirmed by the numerical
results. The values of $G(\lambda)$ can be found in \cite{PRE}.

Also shown in the same figure are the mono-fractal approximation
($\zeta_q = q b/\lambda_0$), which appears to be accurate for
small $q$, and the function
\begin{equation}
\zeta_q = \sqrt{ \left ({\lambda_0 \over \Delta} \right )^2 + {2qb
\over \Delta}} - {\lambda_0 \over \Delta}.
\label{Chertkovscaling}
\end{equation}
that results from a parabolic $G(\lambda)$:
\begin{equation}
G(\lambda) = {(\lambda - \lambda_0)^2 \over 2 \Delta} \ .
\label{Gauss}
\end{equation}

This can be thought as the first term in a Taylor expansion around
$\lambda_0$. It gives a good approximation to obtain the small-$q$
scaling exponents. Expression (\ref{Chertkovscaling}) becomes
exact for the Kraichnan flow \cite{Chertkov98}.

\section{Excitable dynamics}
\label{sec:excitable}

Excitable dynamics \cite{murray} refers to the class of dynamical
systems in which one can identify {\sl activator} and {\sl
inhibitor} variables with the following properties: The {\sl
activator} displays some kind of autocatalitic growth behavior,
but the presence of the {\sl inhibitor} controls it so that the
dynamical system has a stable fixed point as unique global
attractor. The essence of the excitability phenomenon is the
presence of a threshold, such that if there is a perturbation
above it the system variables reach the stable fixed point only
after a large excursion in phase space. This behavior usually
appears when the activator has a temporal response much faster
than the inhibitor, which then takes some time before stopping the
growth of the activator.

Many chemical systems behave in this way, including the famous
Belousov-Zhabotinsky reaction in adequate concentration ranges, or
the electrochemical reactions occurring in nerves and muscles
\cite{murray}. Truscott and Brindley
\cite{Brindley94a,Brindley94b} identified phytoplankton as the
fast activator and zooplankton as the slow inhibitor in models of
the type (\ref{NPZ}), and demonstrated excitable behavior when
Hollings-III grazing functions are used. This opens the
possibility of understanding some plankton dynamics features in
terms of well established scenarios found for excitable chemical
reactions. In particular \cite{Brindley94a} explained features of
red tides in terms of the essentials of excitable dynamics. In
addition, transport processes coupled to excitable dynamics lead
to a rich variety of pattern forming phenomena. The most widely
studied is the appearance of excitable waves, of linear, circular,
or spiral shape, in excitable media with diffusion\cite{murray}.

In the next subsection we summarize the results of
\cite{Brindley97} concerning the generation of plankton patchiness
(via plankton blooms) in excitable media without transport
processes, and in subsection \ref{subsec:ExcitTrans} we will
report some recent results for excitable dynamics under chaotic
advection\cite{PRLexcitable,GRL,NEWexcitable}.

\subsection{Localized blooms from excitable dynamics}
\label{subsec:ExcitNoTrans}

Excitability is a threshold phenomenon: whenever it is crossed,
the system undergoes a large excursion in state space, after which
it returns back to the unexcited state. In an extended system,
different regions may be in different phases of the excitation
cycle. If transport processes are not efficient enough to couple
different regions, excitation will occur only in the places where
the excitation threshold has been reached, whereas the rest of the
system will remain essentially at the equilibrium concentrations.
It was realized in \cite{Brindley97} that this simple mechanism
leads to localized blooms in plankton models. For adequate
parameter values, the temporal evolution of the bloom has the
characteristics of red tides\cite{Brindley94a}.

\begin{center}
\begin{figure}[t]
\epsfig{file=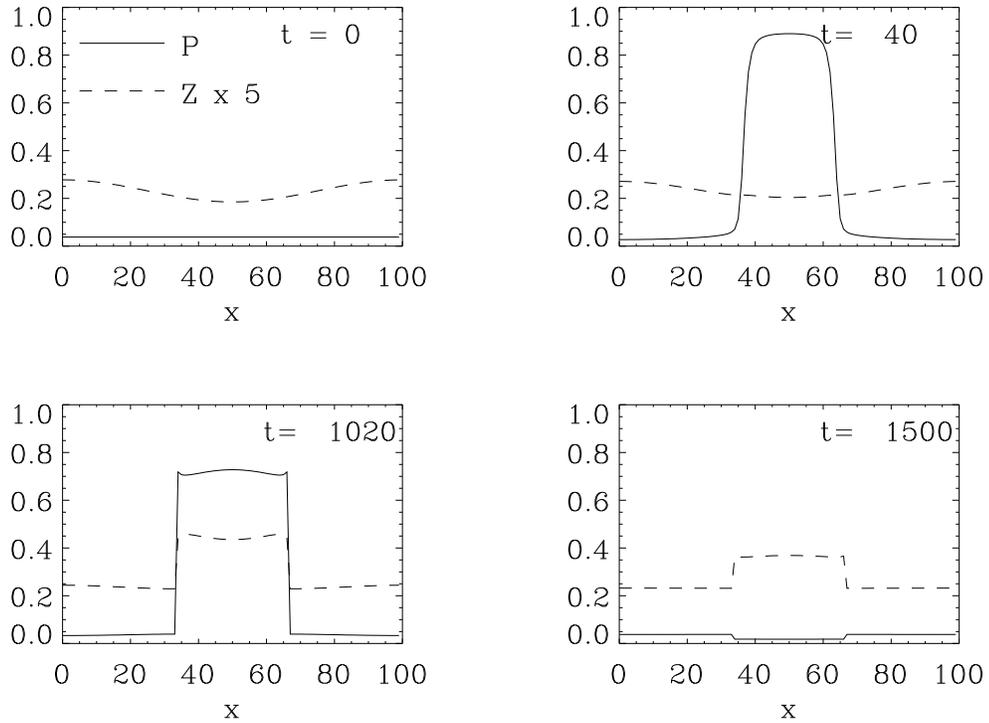,width=.9\linewidth}
\nopagebreak
 \caption{\protect\footnotesize Four stages of a plankton bloom obtained from model
(\ref{BrindleyModel}) starting from the initial condition in the
first panel.}
\label{fig:PlanktonExcitNotrans}
\end{figure}
\end{center}

We illustrate this phenomenon with the model presented in
\cite{Brindley94a}. It is of the form (\ref{NPZ}) but reduced to
two species ($P$-$Z$) with phytoplankton logistic growth and
Hollings-III grazing. In convenient adimensional units it reads
\BA
F_P(P,Z) &=& \alpha P (1-P) -{P^2 \over P^2+P_e^2} Z  \nonumber \\
F_Z(P,Z) &=& \gamma {P^2 \over P^2+P_e^2} Z - m Z \ \ .
\label{BrindleyModel}
\EA
It presents excitable behavior in a range of parameters. In
particular we show in Fig.~\ref{fig:PlanktonExcitNotrans} four
stages of the evolution for $\alpha=0.43$, $P_0=0.053$,
$\gamma=0.005$, and $m=0.34$. The system is distributed in the
one-dimensional line, but each point evolves with the same
dynamics (\ref{BrindleyModel}) independently of the others. Its
evolution thus depends only on its own initial conditions.

The first panel in Fig.~\ref{fig:PlanktonExcitNotrans} shows this
initial condition: Phytoplankton is everywhere at the equilibrium
fixed-point concentration, but zooplankton presents values somehow
smaller than equilibrium in the central zone. In the region where
this depletion is larger than a threshold value, fast
phytoplankton growth occurs, followed by growth in zooplankton and
subsequent recovery of the equilibrium values. The bloom is
localized since there is neither advection nor diffusion to
communicate excitation to neighboring places. Related behavior
occurs if the initial perturbation consists in a local
phytoplankton increase, or in a local increase of the
phytoplankton growth rate $\alpha$.

\subsection{Excitable dynamics under chaotic stirring}
\label{subsec:ExcitTrans}

Except in the most quiescent situations, fluid flow will disturb
the localized blooms of the previous section. Chaotic advection
will deform the static excited patches into long and thin
filaments subjected to stretching and folding. Reference
\cite{PRLexcitable} studied this process including also a small
diffusion. The chemical dynamics considered was the
FitzHugh-Nagumo system, a model originally developed in the
context on nerve excitation, but it is clear that the different
scenarios found would apply to any excitable chemical or
biological dynamics in a chaotic flow.

Three qualitatively different kinds of behavior appeared. In the
case in which the stirring time-scale (measured for example by the
flow Lyapunov exponent) is much slower than the time scale for the
excitation-deexcitation cycle, a localized perturbation propagates
essentially as a deformed version of the circular waves that would
propagate diffusively in a quiescent medium. In the other extreme
case, when the stirring time-scale is faster than the time scale
for growth of the activator, stretching of the filaments and the
associated lateral contraction occur too fast. The largest
gradients associated to filament narrowing increase the diffusive
flux out of them, and they can not be filled-up with new
excitation fast enough. As a result the stretched filaments become
increasingly diluted and concentrations may fall below the
excitation threshold. At this point the excitation will disappear.
This means that a too vigorous stirring will eliminate a small
localized excitation. Perhaps the most surprising behavior appears
in a range of intermediate stirring speeds, such as the dilution
effect stops the growth of the slow inhibitor, but not of the fast
activator: The activator concentration inside the filament remains
saturated, the lateral thinning becomes stopped at a finite width
by the combined effect of excitation and diffusion, and the
filament length increases continuously as a consequence of chaotic
stretching. After some time, in a closed flow after repeated
folding, the filament will fill-up the whole system. The effect of
chaotic advection has changed the character of an initial
excitation from localized to global.  After the whole system is
excited, deexcitation proceeds as in a homogeneous well-mixed
medium.

\begin{center}
\begin{figure}[t]
\epsfig{file=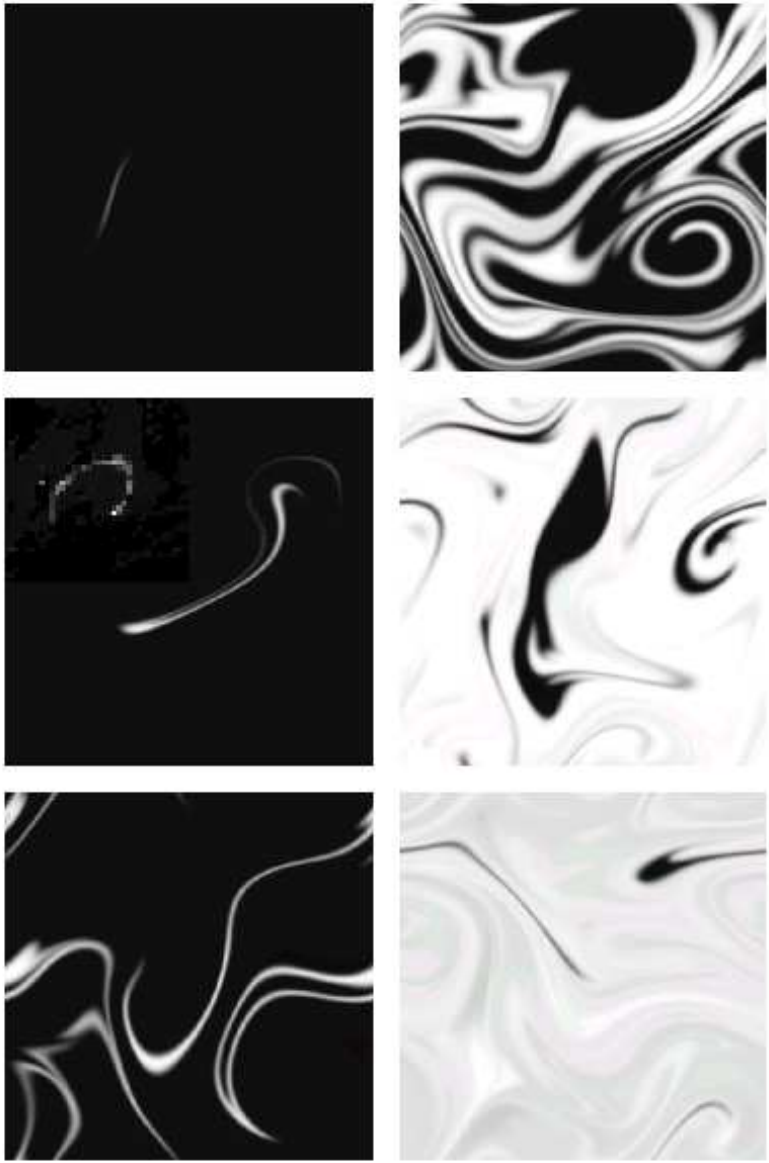,width=.7\linewidth}
\nopagebreak
 \caption{\protect\footnotesize Snapshots of phytoplankton concentration
(darker grey levels indicate smaller concentrations) from an
excitable plankton model. Time runs from top to bottom, and then
from left to right. Periodic boundary conditions are imposed to
have a closed flow. The initial localized excited patch is
stretched and folded until the full system becomes excited. After
this (not shown) homogeneous deexcitation occurs. The inset in the
second panel shows the chlorophyll filament observed from a
satellite sensor (SeaWiFS) 42 days after a fertilization
experiment in the Antartic\protect\cite{NatureFilament} that can
be considered as the seeding of a localized perturbation. }
\label{fig:PlanktonExcitTrans}
\end{figure}
\end{center}

The process is illustrated in Fig.~\ref{fig:PlanktonExcitTrans},
which is obtained \cite{GRL} by integration of
advection-reaction-diffusion equations (\ref{ard}), with the
reaction terms as in the previous subsection (Eq.
(\ref{BrindleyModel})). Advection and diffusion for a nutrient
field is also included (with $F_N=0$). The coupling of the
nutrient with phytoplankton is made via a dependence of the growth
rate $\alpha$ on the nutrient concentration
($\alpha=\alpha_0(1+N/N_0)$). For the flow (closed and
incompressible) a set of randomly seeded eddies has been
used\cite{abraham}. The inset in the second panel of
Fig.~\ref{fig:PlanktonExcitTrans} shows a chlorophyll filament
observed from the satellite sensor SeaWiFS 42 days after a
fertilization experiment in the Antartic\cite{NatureFilament}.
Reference \cite{GRL} interprets this filament as the result of
stretching and folding of the initial excitation introduced in the
fertilization experiment. This would support the interpretation of
the observed dynamics in terms of the concepts of excitability and
chaotic advection. The phenomenon reported here, i.e. the
occurrence of a global excitation by the effect of chaotic
advection on a localized patch, gives a warning about the
possibility of responses of unexpected extension arising from very
localized perturbations.

The essential object in this kind of phenomena is the stretched
filament. In \cite{PRLexcitable,NEWexcitable}, simplified
equations describing the transverse filament profile are
discussed. They are essentially of the form (\ref{martinEq})
adapted to the multicomponent case, and with the simple linear
chemistry term $\mu(t) P$ replaced by the excitable chemistry.
These filament equations can be used to obtain analytical
understanding of the transitions between the different regimes,
and to get quantitative results. For example, the width of the
excited filament is given\cite{PRLexcitable,NEWexcitable} by $w=c
\sqrt{D\alpha}/\lambda$, where $c$ is a numerical constant
depending on the particular excitable model used, $D$ is the
diffusion coefficient, $\alpha$ the phytoplankton linear growth
rate, and $\lambda$ the exponential contraction given by the
strain, that can be identified with the flow Lyapunov exponent.

\section{Some warnings from individual modeling}
\label{sec:individual}

In the present Lectures we have restricted our considerations to
continuous differential equation models describing continuous
concentration variables. The use of maps \cite{chaos2001} is an
alternative to the use of continuous differential equations. But
there is an aspect of real chemical or biological interacting
systems that is missed in both approaches: the discrete nature of
the interacting entities (chemical molecules or the individual
organisms). On general grounds, one would expect just a more noisy
dynamics in models in which individuals are resolved. There are
cases however in which individual quantization produces more
profound effects. We summarize here the results of two recent
references \cite{ShnerbPNAS,YoungNature} that highlight this fact
in very simple examples.

Reference \cite{ShnerbPNAS} discusses the following chemical
scheme:
\BA
A+C &\rightarrow& 2C  \nonumber
\\ C &\rightarrow& 0
\label{PNASsystem}
\EA
  The first autocatalitic equation can be interpreted as a simple
process in which the $C$ particle eats `food' $A$ to
self-replicate (the amount of $A$ is kept constant), and the
second represents the death of $C$. The continuous description in
terms of a reaction-diffusion equation for the concentration $C$
would be
\BE
{\partial C \over \partial t} = (\lambda A - \mu) C + D \nabla^2 C
\label{PNAScontinuum}
\EE
$\lambda$ and $\mu$ are the rates of the first and second reaction
in (\ref{PNASsystem}) respectively. The continuous model predicts
extinction of $C$ if the effective growth rate $\lambda A - \mu$
is negative.

Alternatively one can model the process by a large number of
random walkers, each representing a molecule of $A$ or $C$, that
undergo the transformations (\ref{PNASsystem}) following some
Monte Carlo dynamics. It happens \cite{ShnerbPNAS} that if the
effective growth rate is not too negative, the amount of $C$
increases instead of decreasing. The consideration of the discrete
nature of the individuals has changed death to life. What happens
is that the $C$ individuals die in most of space, but they survive
if close enough to `food' particles. The effect is a diverging
clustering of $C$ particles around $A$ random walkers. This strong
inhomogeneity is missed in the continuous model, that gives the
wrong prediction. The root of the problem is in the use of mean
multiplicative growth rates in (\ref{PNAScontinuum}), whereas in
fact they vary randomly in space and time, reflecting the
positions of the $A$ particles. It is known that mild randomness
in the multiplicative growth factors in linear equations such as
(\ref{PNAScontinuum}) give rise to strongly intermittent
distributions\cite{Sokoloff,Zeldovich}. Intermittency reflects
here in the very inhomogeneous clustering of the surviving
particles.

A related phenomenon is observed in \cite{YoungNature}. There, the
chemical scheme is
\BA
C &\rightarrow& 2C  \nonumber
\\ C &\rightarrow& 0
\label{Naturesystem}
\EA
which leads to essentially the same continuous description
(\ref{PNAScontinuum}), with $\lambda A$ replaced by $\lambda$.
Here again growth is observed in discrete simulations when the
continuous model indicates extinction. Here there are no `food'
particles to cluster around, but clustering occurs around the
initially seeded $C$ particles. There are position correlations
between a parent particle and its descendants (the ones appearing
via the first autocatalitic reaction in (\ref{Naturesystem})),
that are completely neglected in the continuous model.

The authors of \cite{YoungNature} check that the reproductive
correlations persist in the presence of chaotic advection: the
clusters become filaments, but the anomalous survival remains. The
authors are able to describe the correct behavior via continuous
equations, but not for the average concentration, but for binary
correlations. Whether or not a given discrete process can be
modeled in a continuum model, and up to which level of statistics
one should retain randomness and correlations, is a major question
in reacting particle models and in general ecosystem dynamics.

\section{Conclusions}
\label{sec:conclusions}

We have presented a number of processes, inspired by concepts in
Nonlinear Dynamics such as chaos and excitability, that can be
useful to understand generic behaviors in chemical or biological
systems in fluid flows.

For the case of the linearly decaying tracer, a rather complete
description is available. The scaling exponents of the set of
structure functions, which display anomalous (multifractal)
behavior, can be expressed in terms of the decay-time constant and
the probability distribution of the Lagrangian finite-time
Lyapunov exponents of the flow. The decay-time constant makes the
tracer power spectrum steeper than the Batchelor law obtained for
the passive tracer.  The differences between open and closed flows
are important, and the relevance of these results for nonlinear
plankton models has been be pointed out.

The study of excitable-type reactions is motivated by the presence
of excitability in some plankton dynamics models. It contains the
essentials of the observed characteristics of plankton blooms.
Important changes in the response of excitable models are found
when they are placed in a chaotic fluid flow. This may be relevant
for the outcome of ocean fertilization experiments.

Finally, some warnings have been given about the difficulties in
modeling discrete individuals (such as planktonic organisms) in
terms of continuous concentration fields. Aggregation behavior,
reproductive correlations, and other phenomena turn out to be
difficult to introduce in standard continuous models, whereas they
are quite naturally present in individual-based descriptions.

\acknowledgements
We acknowledge Susana Agust\'{\i} for providing us the data shown
in Fig.\ref{fig:transect}, Adrian P. Martin for motivating our
interest in some of the topics presented here, and Carlos Duarte,
Veronique Gar\c{c}on, Peter H. Haynes, Oreste Piro, Tam\'{a}s
T\'{e}l, and Angelo Vulpiani for illuminating discussions. We
thank also the organizers of the 2001 International Summer School
on Atmospheric and Oceanic Sciences (ISSAOS 2001) for the
excellent opportunity to discuss the subject of Chaos in
Geophysical Flows. Financial support from MCyT (Spain) projects
IMAGEN REN2001-0802-C02-01/MAR and CONOCE BFM2000-1108 is greatly
acknowledged. C.L. acknowledges support from MECD (Spain).


\begin{references}
\baselineskip 15pt
{\footnotesize
\bibitem{karoly99} G. K\'{a}roly, A. P\'{e}ntek, Z. Toroczkai, T.
T\'{e}l, C. Grebogi, {\sl Chemical or biological activity in open
flows}, Phys. Rev. {\bf E 59}, 5468 (1999).

\bibitem{edouard} S. Edouard, B. Legras, F. Lef\`{e}vre, R.
Aymard, {\sl The effect of small-scale inhomogeneities on ozone
depletion in the Artic}, Nature {\bf 384}, 444 (1996).

\bibitem{mahadevan} A. Mahadevan, D. Archer, {\sl Modeling the impact
of fronts and mesoscale circulation on the nutrient supply and
biogeochemistry of the upper ocean}, J. Geophys. Res. C {\bf 105},
1209 (2000).

\bibitem{PowellOkubo} T.M. Powell, A. Okubo, {\sl Turbulence, diffusion
and patchiness in the sea}, Phil. Trans. R. Soc. Lond. B {\bf
343}, 11 (1994).

\bibitem{Abraham2} E.R. Abraham and E.M. Bowen,
{\sl Chaotic stirring by a mesoscale surface-ocean flow}, arXiv
preprint {\tt nlin.CD/0204011} (available from {\tt
http://arXiv.org}).

\bibitem{catalan} J. Catalan, {\sl Small-scale hydrodynamics as a
framework for plankton evolution}, Jpn. J. Limnol. {\bf 60}, 469
(1999).

\bibitem{BarnesHughes} R.S.K. Barnes, R.N. Hughes, {\sl An introduction to
marine ecology} (Blackwell Scientific Publications, Boston, 1988).

\bibitem{MannLazier} K.H. Mann, J.R.N. Lazier, {\sl Dynamics of marine ecosystems.
Biological-physical interactions in the oceans} (Blackwell
Scientific Publications, Boston, 1991).

\bibitem{mackas} D.L. Mackas, K.L. Denman, and M.R. Abbott, {\sl Plankton
patchiness: biology in the physical vernacular}, Bull. Mar. Sci.
{\bf 37}, 652 (1985).

\bibitem{kiSs} J.G. Skellam, {\sl Random dispersal in theoretical
populations}, Biometrika {\bf 38}, 196 (1951).

\bibitem{KIsS} H. Kierstead and J.B. Slobodkin, {\sl The size of water masses
containing plankton blooms}, J. Mar. Res. {\bf 12} 141 (1953).

\bibitem{seawifs} Public sites with satellite images that illustrate the phenomena
mentioned here are, for example, the SeaWiFS project home page
({\tt http://seawifs.gsfc.nasa.gov/SEAWIFS.html}) and the Marine
Environment Unit of the Joint Research Centre of the European
Commission ({\tt http://www.me.sai.jrc.it/me-website/}).

\bibitem{timevar} R.E. Wilson, A. Okubo, W.E. Esaias, {\sl A note on
time-dependent spectra for chlorophyll variance}, J. Mar. Res.
{\bf 37}, 485 (1979).

\bibitem{DenmanPlatt} K.L. Denman, T. Platt, {\sl The variance spectrum
of phytoplankton in a turbulent ocean}, J. Mar. Res. {\bf 34}, 593
(1976).

\bibitem{agusti} Discussion of the methodology and of data
collected in a Northern part of the same campaign are in S.
Agusti, C.M. Duarte, D. Vaqu\'{e}, M. Hein, J.M. Gasol, M. Vidal,
{\sl Food-web structure and elemental (C, N and P) fluxes in the
eastern tropical North Atlantic}, Deep Sea Res. II {\bf 48}, 2295
(2001).

\bibitem{scales1} S. Levin, {\sl The problem of pattern and scale in
ecology}, Ecology {\bf 73}, 1943 (1992).

\bibitem{scales2} D.A. Rand, H.B. Wilson, {\sl Using spatio-temporal chaos and
intermediate-scale determinism to quantify spatially extended
ecosystems}, Proc. Roy. Soc. Lond. B {\bf 259}, 111 (1995).

\bibitem{Brindley94b} J.E. Truscott, J. Brindley, {\sl Equilibria,
stability and excitability in a general class of plankton
population models}, Phil. Trans. R. Soc. Lond. A {\bf 347}, 703
(1994).

\bibitem{Edwards01} A.M. Edwards, M.A. Bees, {\sl Generic dynamics
of a simple plankton population model with a non-integer exponent
of closure}, Chaos, Solitons and Fractals {\bf 12}, 289 (2001).

\bibitem{murray} J.D. Murray, {\sl Mathematical biology}
(Springer-Verlag, Berlin, 1993).

\bibitem{Hopf} G.F. Fussmann, S.P. Ellner, K.W. Shertzer, N.G.
Hairstor Jr., {\sl Crossing the Hopf bifurcation in a live
predator-prey system}, Science {\bf 290}, 1358 (2000).

\bibitem{okubo} A. Okubo, {\sl Oceanic diffusion diagrams},
Deep-Sea Res. {\bf 18}, 789 (1971).

\bibitem{martin} A.P. Martin, {\sl On filament width in oceanic
plankton distributions}, J. Plank. Res. {\bf 22}, 597 (2000).

\bibitem{Gollub} G. Voth, G. Haller, J. P. Gollub, {\sl Precision
Measurements of Stretching and Compression in Fluid Mixing}, arXiv
preprint {\tt nlin.CD/0109006} (available from {\tt
http://arXiv.org}).

\bibitem{NatureFilament} E.R. Abraham, C.S. Law, P.W. Boyd, S.J.
Lavender, M.T. Maldonado, and A.R. Bowie, {\sl Importance of
stirring in the development of an iron-fertilized phytoplankton
bloom}, Nature {\bf 407}, 727 (2000).

\bibitem{turing} S.A. Levin, L.A. Segel, {\sl Hypothesis for
origin of plankton patchiness}, Nature {\bf 259}, 659 (1976).

\bibitem{Brindley97} L. Matthews, J. Brindley, {\sl Patchiness in plankton
populations}, Dyn. Stab. Sys. {\bf 12}, 39 (1997).

\bibitem{fede} F. Bartomeus, D. Alonso, J. Catalan, {\sl
Self-organized spatial structures in a ratio-dependent
predator-prey model}, Physica A {\bf 295}, 53 (2001); D. Alonso,
F. Bartomeus, J. Catalan, {\sl Mutual interference between
predators can give rise to Turing spatial patterns}, Ecology {\bf
83}, 28 (2002).

\bibitem{rovinsky} A.B. Rovinsky, H. Adiwidjaja, V.Z. Yakhnin, M.
Mezinger, {\sl Patchiness and enhancement of productivity in
plankton ecosystems due to differential advection of predator and
prey}, OIKOS {\bf 78}, 111 (1997).

\bibitem{abraham}
E. R. Abraham, {\it The generation of plankton patchiness by
turbulent stirring}, Nature {\bf 391}, 577 (1998).

\bibitem{aref} H. Aref, {\sl Stirring by chaotic advection}, J.
Fluid Mech. {\bf 143}, 1 (1984); for an historical account of the
concept of chaotic advection see H. Aref, {\sl The development of
chaotic advection}, Phys. Fluids {\bf 14}, 1315 (2002).

\bibitem{Bohr} T. Bohr, M. Jensen, G. Paladin and A. Vulpiani,
{\sl Dynamical Systems Approach to Turbulence} (Cambridge Univ.
Press, Cambridge, 1998).

\bibitem{Ott} E. Ott, {\sl Chaos in Dynamical Systems} (Cambridge
Univ. Press, Cambridge, 1993).

\bibitem{Falko} G. Falkovich, K. Gaw\c{e}dzki, M. Vergassola, {\sl Particles
and fields in fluid turbulence}, Rev. Mod. Phys. {\bf 73}, 913
(2001).

\bibitem{batchelor} G.K. Batchelor, {\sl Small scale variation of
convected quantities like temperature in turbulent fluid. Part 1.
General discussion and the case of small conductivity}, J. Fluid
Mech. {\bf 5}, 113 (1959).

\bibitem{dissipation} R.K. Dewey, J.N. Moum, {\sl Enhancement of
Fronts by Vertical Mixing}, J. Geophys. Res. {\bf 95}, 9433
(1990).

\bibitem{peter} P.H. Haynes {\sl Transport, stirring and mixing in the
atmosphere}, in {\sl Mixing - Chaos and turbulence}, Ed. by H.
Chate, E. Villermaux and J.M. Chomaz (Kluwer, Dordretch, 1999).

\bibitem{corrsin} S. Corrsin, {\sl The reactant concentration spectrum in turbulent
mixing with a first-order reacion}, J. Fluid Mech. {\bf 11}, 407
(1961).

\bibitem{PRL} Z. Neufeld, C. L\'opez and P.H. Haynes, {\sl Smooth-Filamental Transition
of Active Tracer Fields Stirred by Chaotic Advection}, Phys. Rev.
Lett. {\bf 82}, 2606 (1999).

\bibitem{PRE} Z. Neufeld, C. L\'{o}pez, E. Hern\'andez-Garc\'{\i}a, T. T\'{e}l,
{\sl The multifractal structure of chaotically advected chemical
fields}, Phys. Rev. {\bf E 61}, 3857 (2000).

\bibitem{Chertkov98} M. Chertkov, {\sl On how a joint interaction of two innocent
partners (smooth advection and linear damping) produces a strong
intermittency}, Phys. Fluids {\bf 10}, 3017 (1998).

\bibitem{CHAOS} E. Hern\'{a}ndez-Garc\'{\i}a, Crist\'{o}bal L\'{o}pez, Z.
Neufeld, {\sl Small-scale structure of nonlinearly interacting
species advected by chaotic flows}, CHAOS {\bf 12} (in press,
2002).

\bibitem{Chertkov99} M. Chertkov, {\sl Passive advection in nonlinear medium},
Phys. Fluids {\bf 11}, 2257 (1999).

\bibitem{PCE} C. L\'{o}pez, Z. Neufeld, E. Hern\'{a}ndez-Garc\'{\i}a, P.H. Haynes,
{\sl Chaotic advection of reacting substances: Plankton dynamics
on a meandering jet}, Phys. Chem. Earth {\bf B 26}, 313 (2001).

\bibitem{bower}
A.S. Bower, {\it A simple kinematic mechanism for mixing fluid
parcels across a meandering jet}, J. Phys. Oceanogr. {\bf 21}, 173
(1991).

\bibitem{Tel} T. T\'el,
in {\sl Directions in Chaos}, Vol. 3, Ed. by Hao Bai-Lin (World
Scientific, Singapore, 1990).

\bibitem{Jung} C. Jung, T. T\'el and E. Ziemniak, {\sl Application of scattering chaos
to particle transport in a hydrodynamical flow}, CHAOS {\bf 3},
555 (1993).

\bibitem{Ziemniak} E. Ziemniak, C. Jung and T. T\'el, {\sl Tracer
dynamics in open hydrodynamical flows as chaotic scattering},
Physica {\bf D 76}, 123 (1994).

\bibitem{Sommerer} J.C. Sommerer, H.-C. Ku and H.E. Gilreath, {\sl
Experimental Evidence for Chaotic Scattering in a Fluid Wake},
Phys. Rev. Lett. {\bf 77}, 5055 (1996).

\bibitem{Pascual} M. Pascual, F.A. Ascioti, H. Caswell, {\sl Intermittency
in the plankton: a multifractal analysis of zooplankton biomass
variability}, J. Plank. Res. {\bf 17}, 1209 (1995).

\bibitem{Seuront}
L. Seuront, F. Schmitt, Y. Lagadeuc, D. Schertzer, S. Lovejoy, S.
Frontier, {\sl Multifractal analysis of phytoplankton biomass and
temperature in the ocean}, Geophys. Res. Lett. {\bf 23}, 3591
(1996).

\bibitem{Brindley94a} J.E. Truscott, J. Brindley, {\sl Ocean
plankton populations as excitable media}, Bull. Math. Biol. {\bf
56}, 981 (1994).

\bibitem{PRLexcitable} Z. Neufeld, {\sl Excitable media in a chaotic
flow}, Phys. Rev. Lett. {\bf 87}, 108301 (2001).

\bibitem{NEWexcitable} Z. Neufeld, C. L\'{o}pez, E.
Hern\'{a}ndez-Garc\'{\i}a, O. Piro, {\sl Open and closed excitable
flows}, preprint (2002).

\bibitem{GRL} Z. Neufeld, P. H. Haynes, V. Gar\c{c}on, J. Sudre, {\sl Ocean
fertilization experiments may initiate a large scale phytoplankton
bloom}, Geophys, Res. Lett. (in press, 2002).

\bibitem{chaos2001} C. L\'{o}pez, E. Hern\'{a}ndez-Garc\'{\i}a,
O. Piro, A. Vulpiani, and E. Zambianchi, {\sl Population dynamics
advected by chaotic flows: A discrete-time map approach}, CHAOS
{\bf 11}, 397-403 (2001).

\bibitem{ShnerbPNAS} N. M. Shnerb, Y. Louzoun, E. Bettelheim, and S. Solomon,
{\sl The importance of being discrete: Life always wins on the
surface}, Proc. Nat. Acad. Sci. {\bf 97}, 10322 (2000).

\bibitem{YoungNature} W.R. Young, A.J. Roberts, G. Stuhne, {\sl Reproductive
pair correlations, Brownian bugs and plankton patches}, Nature
{\bf 412}, 328 (2001).

\bibitem{Sokoloff} We thank D.D. Sokoloff for pointing to us this
connection.

\bibitem{Zeldovich} Ya. B. Zeldovich, A.A. Ruzmaikin, D.D.
Sokoloff, {\sl The Almighty Chance} (World Scientific, Singapore,
1990). }
\end{references}
\end{document}